\documentclass{article}

\usepackage{arxiv}

\usepackage[utf8]{inputenc} % allow utf-8 input
\usepackage[T1]{fontenc}    % use 8-bit T1 fonts
\usepackage{hyperref}       % hyperlinks
\usepackage{url}            % simple URL typesetting
\usepackage{booktabs}       % professional-quality tables
\usepackage{amsfonts}       % blackboard math symbols
\usepackage{nicefrac}       % compact symbols for 1/2, etc.
\usepackage{microtype}      % microtypography
\usepackage{lipsum}
\usepackage{caption}
\usepackage{amsmath}
\usepackage{amsthm}
\usepackage{bbold}
\usepackage{siunitx}
\usepackage{longtable}
\usepackage{array}
\usepackage{multirow}
\usepackage{wrapfig}
\usepackage{float}
\usepackage{chapterbib}
\usepackage{colortbl}
\usepackage{pdflscape}
\usepackage{tabu}
\usepackage{threeparttable}
\usepackage{threeparttablex}
\usepackage[normalem]{ulem}
\usepackage{makecell}
\usepackage{xcolor}
\usepackage[round]{natbib}
\usepackage{enumerate}
\usepackage[shortlabels]{enumitem}
\usepackage{pgfplots}
\usepackage{pdflscape}
\newtheorem{theorem}{Theorem}[section]
\newtheorem{lemma}{Lemma}
\newtheorem{assumption}{Assumption}

\title{Causal machine learning for high-dimensional mediation analysis using interventional effects mapped to a target trial}

\author{
   Tong Chen \\
   Clinical Epidemiology and Biostatistics Unit, 
   Murdoch Children’s Research Institute\\
   Melbourne Dental School,
   University of Melbourne\\
   \AND
   Stijn Vansteelandt \\
   Department of Applied Mathematics, Computer Science and Statistics,
   Ghent University \\
   \AND
   David Burgner \\
   Inflammatory Origins,
   Murdoch Children’s Research Institute\\
   Department of Paediatrics,
   University of Melbourne\\
   \And
   Toby Mansell \\
   Inflammatory Origins,
   Murdoch Children’s Research Institute\\
   Department of Paediatrics,
   University of Melbourne\\
   \And
   Margarita Moreno-Betancur \\
   Clinical Epidemiology and Biostatistics Unit,  
   Murdoch Children’s Research Institute\\
   Department of Paediatrics,
   University of Melbourne\\
}

\pgfplotsset{compat=1.18} 
\begin{document}
\maketitle

\begin{abstract}
Causal mediation analysis examines causal pathways linking exposures to disease. The estimation of interventional effects, which are mediation estimands that overcome certain identifiability problems of natural effects, has been advanced through causal machine learning methods, particularly for high-dimensional mediators. Recently, it has been proposed interventional effects can be defined in each study by mapping to a target trial assessing specific hypothetical mediator interventions. This provides an appealing framework to directly address real-world research questions about the extent to which such interventions might mitigate an increased disease risk in the exposed. However, existing estimators for interventional effects mapped to a target trial rely on singly-robust parametric approaches, limiting their applicability in high-dimensional settings. Building upon recent developments in causal machine learning for interventional effects, we address this gap by developing causal machine learning estimators for three interventional effect estimands, defined by target trials assessing hypothetical interventions inducing distinct shifts in joint mediator distributions. These estimands are motivated by a case study within the Longitudinal Study of Australian Children, used for illustration, which assessed how intervening on high inflammatory burden and other non-inflammatory adverse metabolomic markers might mitigate the adverse causal effect of overweight or obesity on high blood pressure in adolescence. We develop one-step and (partial) targeted minimum loss-based estimators based on efficient influence functions of those estimands, demonstrating they are root-n consistent, efficient, and multiply robust under certain conditions.

\end{abstract}

% keywords can be removed
\keywords{mediation, target trial, interventional effects, causal machine learning, TML, multiply robust}

\section{Introduction}
\label{sec1}

In clinical and public health research, understanding the causal biological pathways that link exposures and disease risk can inform the development of interventions on intermediate pathways to reduce disease risk in the exposed. Longitudinal cohort studies are increasingly able to generate rich biomarker data (particularly omics data), creating new opportunities to investigate these potential pathway interventions. For example, in the population-derived cohort study that motivated this work, the Longitudinal Study of Australian Children (LSAC), a key research objective was to examine the extent to which intervening on inflammatory and other metabolic pathways could mitigate the increased risk of high blood pressure resulting from a higher body mass index (BMI) in early adolescence.

Recent developments in causal mediation analysis methods aim to delineate the causal pathways linking an exposure and an outcome. Specifically, these methods aim to quantify the extent to which the causal effect of an exposure on an outcome is "mediated" by one or more intermediate variables, known as mediators \citep{baron1986moderator}. Mediation methods have emphasised the decomposition of the causal effect into so-called natural direct and indirect effects defined in the potential outcomes framework \citep{neyman1923, rubin1974estimating}. Specifically, in the single-mediator setting, they are defined based on hypothetical individual-level interventions where each individual is set to be exposed, while the mediator is set to the level it would have been without exposure. Such intervention studies are not achievable in the real world, so these effects may be less informative for studies aimed at informing pathway interventions. Further, the conditions necessary for identifying natural direct and indirect effects have been extensively explored and critiqued \citep{robins1992identifiability,pearl2001direct}. In particular, it has been noted that the identification of natural effects relies on assumptions concerning cross-world counterfactuals that cannot be guaranteed to be satisfied even in a randomised controlled trial \citep{robins2010alternative}. Additionally, natural effects, as originally defined, cannot be identified in the presence of exposure-induced mediator-outcome confounders \citep{avin2005identifiability, TchetgenTchetgen&VanderWeele2014}, also known as intermediate confounders. However, in the context of multiple mediators, certain path-specific natural effects, defined in terms of cross-world counterfactuals, can still be identified and may be of substantive interest \citep{VanderWeele2014Mediation, vansteelandt2017interventional}.

To address these limitations, \cite{vanderweele2014effect} and \cite{vansteelandt2017interventional} introduced interventional effects. In the single-mediator setting, these effects are defined based on interventions where each individual is set to be exposed, while the mediator is set to a random draw from the counterfactual distribution of the mediator under no exposure (possibly given covariate values). Interventional effects can be identified without assumptions regarding cross-world counterfactuals and in settings with intermediate confounders. More importantly, these effects can be interpreted as the effects of a hypothetical intervention setting the exposure and shifting the mediator distribution under exposure to the level under no exposure \citep{moreno2018understanding}. Although interventions achieving such shifts may not always be the most plausible, interventional effects still provide a valuable avenue to inform potential intervention targets in many settings where actual interventions are not yet available or cannot be studied for the outcomes of interest \citep{Moreno-Betancur2021mediation}. Such is the case in the aforementioned motivating example concerning potential interventions on inflammatory and other metabolic pathways. Further, \cite{Moreno-Betancur2021mediation} considered the setting with multiple mediators, and proposed defining interventional effects explicitly in terms of a "target trial" \citep{hernan2016using} where treatment strategies are specified to reflect the hypothetical interventions of interest, which are encoded by shifts in the joint distribution of mediators. This approach results in target estimands that more directly address the research questions of interest, acknowledging the ultimate interventional intent of many studies examining mediation questions, as has been illustrated in several published epidemiological studies \citep{Dashti2022,Goldfeld2023Addressing,Afshar2024}.

In their study, \cite{Moreno-Betancur2021mediation} considered a g-computation approach for the estimation of the proposed target estimands. This approach uses Monte Carlo simulation to estimate joint mediator densities via parametric models. Unfortunately, this approach is not feasible in high-dimensional mediation problems, such as with metabolite data, where the number of mediators is large relative to the number of study participants. In high-dimensional mediation problems, traditional regression-based methods produce biased estimates with large variances due to the curse of dimensionality \citep{donoho2000high} and collinearity \citep{fan2010selective}. In this context, it is natural to question whether predictive machine learning algorithms could replace parametric models within the g-computation approach, as they are well-suited for handling high-dimensional data and do not need to assume a specific functional form. However, directly applying machine learning methods within g-computation can result in biased point estimates and invalid confidence intervals due to their slower convergence rates \citep{van2006targeted}.

An effective solution to this issue is to use causal machine learning estimators, which can incorporate doubly robust methods such as targeted minimum loss-based (TML, \citep{van2006targeted}) estimation and double/debiased machine learning (DML, \citep{chernozhukov2018double}) when combined with machine learning algorithms. These methods, which have been developed for several estimands including average causal effects, incorporate additional debiasing procedures that enable the use of machine learning methods to obtain valid statistical inference. Recently, several causal mediation methods based on causal machine learning estimators have been developed, including for interventional effects \citep{Benkeser2021nonpara, Díaz2021nonpara, Farbmacher2022Causal, Rudolph2024Practical, ran2024nonparametricmotioncontrolfunctional, liu2024generaltargetedmachinelearning}. These estimators can incorporate machine learning methods and handle high-dimensional mediators. Additionally, they attain the nonparametric efficiency bound under certain conditions and maintain multiple robustness, meaning these estimators remain consistent when some, but not necessarily all nuisance parameters (i.e., the parameters that are not of direct interest but are necessary for estimating the target parameter) are consistently estimated. In this work, we develop causal machine learning estimators for interventional effects in the context of multiple (possibly high-dimensional) mediators, for estimands that are defined through mapping them to a target trial, thereby addressing a significant gap given the relevance of these effects for real-world studies. 

We focus on three target estimands, corresponding to interventional effects that map to two different hypothetical mediator interventions that shift the joint mediator distribution. Specifically, we consider estimands that map to (1) an intervention that shifts the distribution of a single mediator while accounting for the flow-on effects on its causal descendants, (2) the same intervention, shifting the distribution of a single mediator, but without accounting for flow-on effects (for settings where the causal ordering of mediators is unknown), and (3) an intervention that shifts the joint distribution of all mediators. As shown by \cite{Moreno-Betancur2021mediation}, these estimands are identifiable under a set of causal assumptions, including an assumption regarding the hypothetical interventions (no causal effect on the outcome other than through mediator distributional shifts) and an exchangeability assumption (no residual exposure-outcome or mediator-outcome confounding given measured baseline confounders). Measured intermediate (exposure-induced mediator-outcome) confounders are allowed, by treating them as other mediators in defining the effects. Under these assumptions, for each estimand we derive causal machine learning estimators based on the efficient influence function and show that these estimators are root-n consistent, multiply robust and efficient under certain (rate) conditions \citep{van2011targeted, Díaz2021nonpara}. Additionally, we illustrate the developed methods in an analysis of the aforementioned study within the LSAC cohort that motivated this work.  

Of note, our estimands (2) and (3) and corresponding estimators are equivalent to those proposed by \cite{Díaz2021nonpara, Benkeser2021nonpara} for scenarios with a single mediator and high-dimensional intermediate confounders and with high-dimensional mediators without intermediate confounders, respectively. Estimand (1) had not yet been studied and our derivation of an estimator extends previous developments by \cite{Díaz2021nonpara} and \cite{Rudolph2024Practical}.

The rest of this article is organised as follows. Section 2 introduces the motivating example. In Section 3, we introduce notation and define three interventional effects estimands of interest by mapping them to target trials. In Section 4, we derive the efficient influence function and causal machine learning estimators for the target estimands defined in Section 3. In Section 5, we apply these estimators to the LSAC study. The discussion is in Section 6. An R package \textit{medoutconRCT} implementing the proposed methods is available at \url{github.com/XXXX}.

\section{Motivating example: Longitudinal Study of Australian Children (LSAC)}\label{mot-exa}

Cardiovascular disease (CVD, heart attack, and stroke) is the leading cause of death worldwide, but it is considered largely preventable through interventions that target both traditional and emerging risk factors earlier in life before the disease occurs \citep{Weintraub2011}. Traditional CVD risk factors include overweight and obesity \citep{Tiffany2021Obesity}, which are associated with adverse cardiovascular measures that are predictors of CVD events in adults, such as increased blood pressure and changes to the microvasculature  \citep{Liu2020}. In childhood and adolescence, circulating levels of metabolites, including markers of inflammation (such as cumulative systemic inflammatory marker glycoprotein acetyls, GlycA) and other cardiometabolic metabolites, are associated with CVD risk \citep{Mansell2022early} and there is emerging evidence that these metabolomic markers partly mediate the relationship between overweight/obesity and adverse cardiovascular measures \citep{Xu2017}. This raises an important research question regarding the extent to which the adverse causal effect of overweight or obesity on high blood pressure, an early marker of cardiovascular disease risk, can be mitigated by hypothetical interventions targeting high inflammatory burden (measured by GlycA) and/or other non-inflammatory adverse metabolomic pathways. Addressing this question is a key step for future intervention development.

To investigate this broad question, we used data from the LSAC B-cohort, an Australian population-based prospective cohort study of children recruited in 2004 aged 0-1 years \citep{Sanson2004GrowingUI}. LSAC has collected health and environmental data on these children through in-home assessments every two years. Between LSAC wave 6 (aged 10-11 years) and wave 7 (aged 12-13 years), a one-off multidimensional physical health and biomarker module known as the Child Health CheckPoint ($n=1874$) was conducted \citep{clifford2019}. This was designed to capture biological data from adolescents including cardiovascular measures, metabolomic markers, and inflammation markers. For our analysis, we drew data from the LSAC waves 6 and 7, along with its interpolated CheckPoint study, including only records without missing data for the analyses, resulting in a sample size of $n = 978$.

In the context of the LSAC study, our specific research questions were: (i) What is the impact on blood pressure of a hypothetical intervention (e.g., a medication) that shifts the distribution of high inflammatory burden, as measured by GlycA, in adolescents with overweight or obesity, to the levels in those without overweight or obesity? and (ii) What is the impact on blood pressure of a hypothetical intervention that shifts the joint distribution of high inflammatory burden and other non-inflammatory adverse metabolomic markers, in adolescents with overweight or obesity, to the levels in those without overweight or obesity?

\begin{figure}[H]
  \centering
  \includegraphics[width=\linewidth]{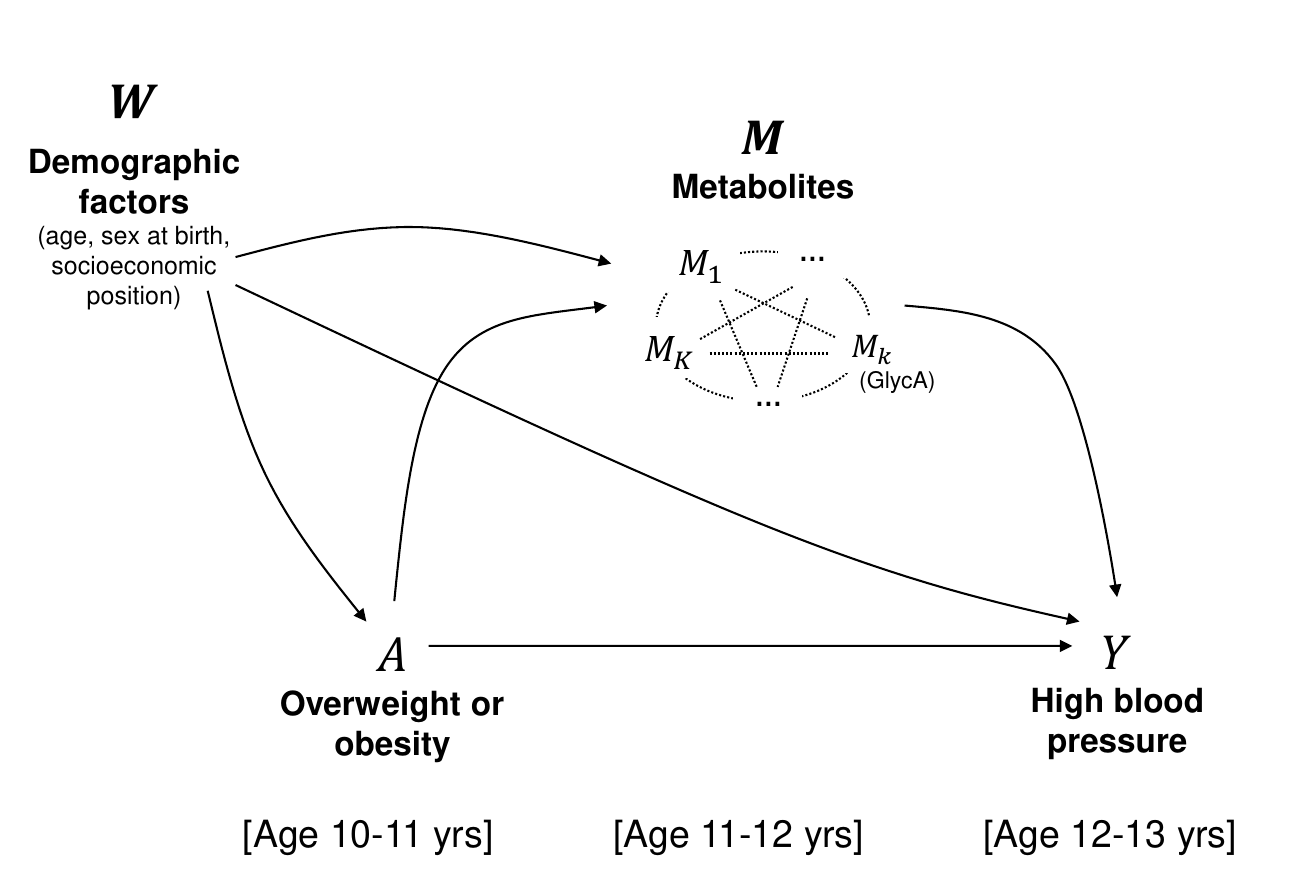}
  \caption{Directed Acyclic Graph (DAG) illustrating the assumed causal structure in the LSAC example. Ellipses (``$\dots$”) represent assumed correlations among mediators. We assume all mediators are correlated but remain agnostic about their causal order unless explicitly specified.}
  \label{dags}
\end{figure}

Specifically, the exposure variable was binary, indicating overweight or obesity status at age 10–11 years, derived by dichotomizing the BMI at 21.3 $\text{kg/m}^2$ (the 85th percentile). The outcome variable was a binary indicator of high blood pressure at age 12–13 years, obtained by dichotomizing average systolic blood pressure at 120 mm Hg (the 95th percentile). Baseline confounders consisted of demographic variables at age 10–11 years, specifically age, sex assigned at birth, and socioeconomic position. The mediators consisted of 70 different metabolites at age 11-12 years, capturing non-inflammatory adverse metabolomic pathways and inflammatory status, the latter defined by dichotomising the GlycA measure. As data on GlycA are relatively sparse in this age group, we explored different dichotomisation cutoffs, specifically at the $\geq 50$th and $\geq 75$th percentiles, to define the indicator of high inflammatory burden. Skewed metabolites were log-transformed as previously described \citep{Ellul106}. A detailed list of these metabolites is provided in Section 6 of the Supplementary Material. The assumed causal structure is depicted in the directed acyclic graph (DAG) in Figure~\ref{dags}.

\section{Defining interventional effects mapped to a target trial}
\label{sec2}
\subsection{Notation}
For illustrative purposes, we define notation in the context of the LSAC example, but note that the definitions and developments would apply to other analogous problems. Let $A \in \{a^\star, a^\prime\}$ denote a binary exposure for overweight or obesity status in adolescence, where $A=a^\prime$ indicates the presence of overweight or obesity, and $A=a^\star$ otherwise. Let $Y$ denote a binary outcome for high blood pressure status, with $Y = 1$ indicating the presence of high blood pressure and $Y=0$ otherwise. Let $M=(M_1, \dots, M_K)$ denote a vector of potentially correlated mediators that may be either continuous or categorical. We denote a particular mediator of interest by $M_k$, which in the LSAC example is an indicator of high inflammatory burden, with $M_k = 1$ indicating the presence of high inflammatory burden and $M_k = 0$ otherwise, while the remaining components of $M$ correspond to measured metabolites (see the list in Section 6 of the Supplementary Material). Let $W$ denote a vector of selected baseline confounders.

Suppose independent and identically distributed data on $O = (Y, A, M, W)$ are collected from $n$ subjects, denoted by $O_1,\dots, O_n$. Moreover, let $P$ denote the distribution function of $O$ which belongs to the nonparametric statistical model $\mathcal{M}$ defined on $O$ and $\mathrm{p}$ denote the corresponding probability density function.

Let $H$ represent a generic hypothetical intervention that shifts the joint mediator distribution (specific examples provided in Section \ref{med-int}), where $H = 1$ indicates that the intervention is received, and $H = 0$ otherwise. Let $Y_a$ denote the potential outcome when $A$ is set to $a$; $Y_{ah}$ denote the potential outcome when we set $A = a$ and $H = h$; $M_{ta}$ denote the status of $t$-th mediator when we set $A = a$ ($t=1,\dots, K)$; and $M_{\cdot a} = (M_{1a}, \dots, M_{Ka})$. 

We further define some nuisance functions. For a random variable $X$, we define $\mathrm{b}(x) = \mathrm{E}(Y|X=x)$ as the true expectation of the outcome given $X=x$, which is referred to as the true outcome model.  Let $\mathrm{g}(a \mid w)$ denote the true probability mass function of $A$ given $W=w$. Similarly, let $\mathrm{q}(m_k \mid a, w)$ represent the true conditional density of $M_k$ given $(A, W) = (a, w)$, and $\mathrm{r}(m_k \mid a, z, w)$ denote the true conditional density of $M_k$ given $(A, Z, W) = (a, z, w)$.

\subsection{Hypothetical mediator interventions} \label{med-int}
Following \cite{Moreno-Betancur2021mediation}, we define interventional effects mapped to a target trial that examines the impact of hypothetical mediator interventions, which shift the joint mediator distribution in specified ways according to the research questions of interest. Specifically, to answer our two specific research questions defined in Section \ref{mot-exa}, we define interventional effects that assess the following hypothetical mediator interventions: 

\subsubsection{Interventions that shift the distribution of the mediator \texorpdfstring{$M_k$}{M\_k}, accounting for flow-on effects on descendant mediators}

To answer research question (i), we focus on a target trial in which one of the treatment strategies is to assign individuals to the exposure ($A=a')$ as well as a hypothetical intervention that shifts the distribution of high inflammatory burden ($M_k$) to the levels in those set to no exposure ($A=a^*)$ given $W$. Per the above, we consider the realistic scenario where all mediators (metabolites) are potentially correlated due to biological mechanisms, and such correlations are partially broken by the hypothetical intervention. Thus, to formally define the shift in the joint mediator distribution resulting from such an intervention, we first consider the setting where a causal ordering among the mediators can be assumed. In that setting, we would expect an intervention that shifts the distribution of GlycA (high inflammatory burden ($M_k$)) to have flow-on effects on the descendant metabolites after the intervention. 
 
Let $Z = (M_{1}, M_{2}, \dots, M_{k-1})$ be the vector of mediators that are causal ancestors of $M_k$, $L= ( M_{k+1},\dots, M_{K}) $ be the vector of mediators that are causal descendants of $M_k$, and $P_M(m)$ denote the joint mediator distribution, where $M = (M_1,\dots,M_K) = (Z, M_k, L)$ and $m = (z,m_k,l)$. As before, let $Z_a = (M_{1a}, M_{2a}, \dots, M_{k-1 a})$ and $L_a= ( M_{k+1 a},\dots, M_{K a})$ denote the potential values for $Z$ and $L$ respectively when setting the exposure $A$. 
 
Then, accounting for flow-on effects on descendant mediators ($L$), the hypothetical intervention of interest for addressing research question (i) would result in the following joint mediator distribution:
\begin{equation*}
P_M(m) = P(Z_{a^{\prime}} = z \mid W) \times P(M_{k a^\star}= m_k | W) \times P(L_{a^\prime} = l | W, Z_{a^{\prime}} = z, M_{k a^\prime}= m_k).
\end{equation*}
In this framework, it is not necessary to establish the full causal ordering of all mediators. Instead, we only need to determine whether a mediator is a causal ancestor, meaning it is a member of the vector $Z$, or a causal descendant, meaning it is a member of the vector $L$, relative to the mediator of interest, $M_k$.
\subsubsection{Interventions that shift the distribution of the mediator \texorpdfstring{$M_k$}{M\_k}, without accounting for flow-on effects (e.g. if causal ordering is unknown)}
We again consider research question (i) but in the setting where it may be difficult to posit a causal ordering of mediators with certainty, as in our example. In such cases, we can also examine the potential impact of hypothetical interventions that would shift the distribution of mediator $M_k$ assuming there are no flow-on effects on other mediators. In this simplified scenario, the vector $Z_a$ denotes the vector $M_{\cdot a}$ excluding $M_{ka}$ and the hypothetical intervention of interest would result in the following joint mediator distribution:
\begin{equation*}
P_M(m) = P(Z_{a^\prime} = z|W) \times P(M_{k a^\star}= m_k |W ) ,  
\end{equation*}
where now $M = (M_1,\dots,M_K) = (Z, M_k)$ and $m = (z,m_k)$.

For interventions (1) and (2), we consider hypothetical interventions that target a single mediator of interest $M_k$ based (conditional) solely on the baseline confounders, without incorporating information from other mediators $Z$ (e.g. medication administered on the basis of available demographic information only). This is why $P(M_{k a^\star}= m_k \mid W)$ is not conditioned on $Z_{a^\star}$ in the above joint distributions.

\subsubsection{Interventions that shift the joint distribution of all mediators}
To answer research question (ii), we focus on a target trial in which one of the treatment strategies is to assign individuals to the exposure ($A=a')$ as well as a hypothetical intervention that shifts the joint distribution of all mediators to the levels in those set to no exposure ($A=a^*)$ given $W$.
This amounts to a hypothetical intervention that could eliminate the adverse metabolic (non-inflammatory) and inflammatory impacts of overweight or obesity. This hypothetical intervention would result in the following joint mediator distribution:
\begin{equation*}
P_M(m) = P(M_{\cdot a^\star} = m \mid W),
\end{equation*}
where $M = (M_1,\dots,M_K)$ and $m = (m_1,\dots,m_K)$. 

\subsection{Target estimand and identification}\label{tar-est}
For a given hypothetical intervention $H$, our target estimand is the interventional indirect effect (IIE), which is defined as $\text{IIE} = E(Y_{a^\prime}) - \theta$, where
\begin{align*}
   \theta=\int E\left(Y_{a^\prime m} \mid W = w\right) \times P_M(m) \times P(W=w) \mathrm{d} m \mathrm{d} w,
\end{align*}
where $P_M (m)$ is the joint mediator distribution following the intervention. Thus, this estimand contrasts the outcome expectation under two treatment strategies (arms) of a target trial: one strategy involves setting $A=a$ and $H=0$, so that the joint mediator distribution is the one that naturally arises when setting $A=a$; and another strategy involves setting $A=a$ and $H=1$, so that the joint mediator distribution has been shifted according to the given hypothetical intervention $H$ \citep{vansteelandt2017interventional, Moreno-Betancur2021mediation}. Identification and estimation of $E(Y_a)$, corresponding to the outcome expectation under the first treatment strategy, have been extensively studied in the literature. We henceforth focus on the identification and estimation of $\theta$, the outcome expectation under the second strategy. 

\cite{Moreno-Betancur2021mediation} showed that the identification results below hold under the following assumptions (I) standard positivity assumptions; (II) no causal effect of the hypothetical intervention $H$ on the outcome other than through mediator distributional shifts; (III) no residual exposure-outcome or mediator-outcome confounding after conditioning on $W$, meaning $Y_{a h} \perp A \mid W$ (given $H$ is the mediator intervention), and no residual exposure-mediator confounding, meaning $M_{\cdot a} \perp A \mid W$; and (IV) the following consistency assumptions: $Y_{ah}=Y$ when $A=a$ and $H=h, \text{ and }M_{ka}=M_{k}$ when $A=a$ for $k=1, \ldots, K$.

\begin{itemize}
    \item Under hypothetical intervention (1), $Z$ represents the causal ancestor of $M_k$, while $L$ represents its causal descendants. The specific target estimand in this case, denoted by $\theta_k^\prime$, can be identified as:
    \begin{align}
        \theta_k^\prime = \int \mathrm{b}(a^\prime, m_k, z, l, w) \mathrm{p}(z \mid a^\prime, w) \mathrm{q}(m_k \mid a^\star,w) \mathrm{p}(l \mid a^\prime, z, m_k, w) \mathrm{p}(w) \mathrm{d}l \mathrm{d}m_k \mathrm{d}z \mathrm{d}w, \label{t1}
    \end{align}
    and the corresponding interventional indirect effect is defined as $\text{IIE}_k^\prime = E(Y_{a^\prime}) - \theta_k^\prime$. 
    \item Under hypothetical intervention (2), $Z$ represents the vector of all mediators excluding $M_k$. The corresponding target estimand can be identified as:
    \begin{align}
        \theta_k = \int \mathrm{b}(a^\prime,m_k, z, w) \mathrm{q}(m_k \mid a^\star,w) \mathrm{p}(z \mid a^\prime, w) \mathrm{p}(w) \mathrm{d}m_k \mathrm{d}z \mathrm{d}w,
    \end{align} 
    and the corresponding interventional indirect effect is defined as $\text{IIE}_k = E(Y_{a^\prime}) - \theta_k$. 
    \item The target estimand under hypothetical intervention (3) can be identified as
    \begin{align}
        \theta_{all} = \int \mathrm{b}(a^\prime,m,w) \mathrm{q}(m \mid a^\star,w) \mathrm{p}(w) \mathrm{d}m \mathrm{d}w,
    \end{align} 
    and the corresponding interventional indirect effect is defined as $\text{IIE}_{all} = E(Y_{a^\prime}) - \theta_{all}$.
\end{itemize}
Note that some of the identification assumptions may not be directly assessable given that $H$ is hypothetical.

\section{Efficient Estimation}
We now develop estimation methods for these target estimands. Given the relationships amongst these, we focus primarily on deriving the efficient estimators for $\theta_k^\prime$ under the nonparametric model, with the results for $\theta_k$ and $\theta_{all}$ provided in the Supplementary Material. Of note, our proposed estimators and implementations for three target estimands are applicable to settings with a binary exposure $A$ and both binary and continuous outcome variable $Y$. For the estimation of estimands $\theta_k^\prime$ and $\theta_k$, it is required that the mediator $M_k$ be binary and there are no restrictions on the types of remaining mediators $(M_1,\dots,M_{k-1},M_{k+1},\dots,M_K)$. For the estimation of $\theta_{all}$, there are no restrictions on the types of mediators $M = (M_1,\dots,M_K)$.

\subsection{Efficient influence function for \texorpdfstring{$\theta_k^\prime$}{theta\_k\_prime}}
In this section, we construct locally efficient estimators for $\theta_k^\prime$ based on the efficient influence function (EIF). They allow the use of flexible machine learning algorithms while ensuring valid statistical inference \citep{pfanzagl1982contributions}.  Additionally, these estimators possess the property of multiple robustness, meaning they remain consistent even if certain components of the data distribution are misspecified and inconsistently estimated, provided that (sufficiently many) others are consistently estimated. We define the indicator function $\mathbb{1}\{a = a^\prime\}$  equals $1$ if $A = a^\prime$ and $0$ otherwise.

\begin{theorem} \label{EIF-theory}
(Efficient influence function for $\theta_k^\prime$) For fixed $a^\prime$ and $a^\star$, we define

\begin{equation*}
\begin{aligned}
s(a,z,w) &=  \int \mathrm{b}(m_k, a, z, l, w) \mathrm{q}(m_k\mid a^\star, w) \mathrm{p}(l\mid a,w,z,m_k)  \mathrm{d}m_k\mathrm{d}l\\
u(a, m_k, w) &=\int \mathrm{b}\left(m_k, a, z, l, w\right) \mathrm{p}\left(z \mid a, w\right) \mathrm{p}\left(l \mid a, w, z, m_k\right) \mathrm{d} z \mathrm{d} l, \\
v(a, w) &=\int \mathrm{b}\left(m_k, a, z, l, w\right) \mathrm{p}(z \mid a, w) \mathrm{p}\left(l \mid a, w, z, m_k\right) \mathrm{q}\left(m_k \mid a^\star, w\right) \mathrm{d} m_k \mathrm{d} z \mathrm{d} l.
\end{aligned}
\end{equation*}

The efficient influence function $D_{P}(o)$ for $\theta_k^\prime$ in a nonparametric model is

\begin{equation}
\begin{aligned} D_{P}(o) &=  D_{P,Y}(o) + D_{P,Z}(o) + D_{P,M_k}(o) + D_{P,L}(o) +  D_{P,W}(o),\text{ where}\\
D_{P,Y}(o) &= \frac{\mathbb{1}\{a = a^\prime\}}{\mathrm{g}\left(a^\prime \mid w\right) } \frac{\mathrm{q}\left(m_k \mid a^\star, w\right)}{\mathrm{r}(m_k \mid a^\prime, z, w)} \left( y - \mathrm{b}(m_k, a^\prime, z, l, w) \right)  \\
D_{P,Z}(o) &= \frac{\mathbb{1}\{a = a^\prime\}}{\mathrm{g}\left(a^\prime \mid w\right) } \left(  s(a,z,w)  - v(a^\prime,w) \right)  \\ 
D_{P,M_k}(o) &= \frac{\mathbb{1}\{a = a^\star\}}{\mathrm{g}\left(a^\star \mid w\right)}\left(u(a^\prime,m_k,w)  -  \int u(a^\prime,m_k,w) \mathrm{q}(m_k \mid a^\star, w) \mathrm{d}m_k \right)  \\
D_{P,L}(o) &= \frac{\mathbb{1}\{a = a^\prime\}}{\mathrm{g}\left(a^\prime \mid w\right) } \frac{\mathrm{q}\left(m_k \mid a^\star, w\right)}{\mathrm{r}(m_k \mid a^\prime, z, w)} \left(\mathrm{b}(m_k, a^\prime, z, l, w) - \int \mathrm{b}(m_k, a^\prime, z, l, w) \mathrm{p}(l \mid a^\prime,w,z,m_k) \mathrm{d}l\right)  \\
D_{P,W}(o) & =  v(a^\prime,w) - \theta_k^\prime
\end{aligned}\label{EIF_prime_eq}
\end{equation} \label{EIF_prime}
\end{theorem}

The proof of Theorem \ref{EIF_prime} is provided in Section 1 of the Supplementary Material. 

\subsection{Efficient influence function for \texorpdfstring{$\theta_k^\prime$}{theta\_k\_prime} - alternative representation for high-dimensional \texorpdfstring{$L$}{L} and \texorpdfstring{$Z$}{Z}}\label{alternative-eif}
In a high-dimensional setting, such as with metabolites in the LSAC example, estimating the high-dimensional densities for $Z$ and $L$ and their associated integrals can be challenging. To overcome the challenge, we work out the following alternative representation of the EIF for the situation where $Z$ and $L$ can be high-dimensional, but $M_k$ is low-dimensional. Define the following conditional density ratio
\begin{equation*}
\begin{aligned}
\mathrm{e}(m_k, a, z, w)=\frac{\mathrm{p}(z \mid a, w)}{\mathrm{p}(z \mid a, m_k, w)} = \frac{\mathrm{q}(m_k \mid a,w)}{\mathrm{r} (m_k\mid a,z,w)}.
\end{aligned}
\end{equation*}
Then the terms in the EIF that involve high-dimensional densities for $L$ and $Z$ can be reformulated as

\begin{align}
s(a,z,w)
&= \mathrm{E} \left( \mathrm{b}\left(m_k, a, z, l, w\right) \frac{\mathrm{q}\left(m_k \mid a^\star, w\right)}{\mathrm{r}(m_k \mid a,z,w)} \mid A=a, Z=z, W=w\right) \label{eqs}\\
u(a, m_k, w)
& = \mathrm{E} \left( \mathrm{b}\left(m_k, a, z, l, w\right) \mathrm{e}(m_k, a, z, w) \mid A=a, M_k = m_k, W=w  \right) \label{equ} \\
v(a,w) &= \mathrm{E} \left( \mathrm{b}\left(m_k, a, z, l, w\right) \frac{\mathrm{q}\left(m_k \mid a^\star, w\right)}{\mathrm{r}(m_k \mid a,z,w)} \mid A=a , W=w\right) \label{eqv}\\
\int \mathrm{b}(m_k, a^\prime, z, l, w)& \mathrm{p}(l \mid a^\prime,w,z,m_k) \mathrm{d}l = \mathrm{E}(\mathrm{b}(m_k, a^\prime, z, l, w) \mid A=a^\prime, Z=z, M_k = m_k, W=w).
\end{align}

Additionally, following the approach of \cite{Díaz2021nonpara}, and in a setting with binary $M_k$ as in the LSAC example, the expression for $D_{P,M_k}(o)$ in Equation (\ref{EIF_prime_eq}) can be further simplified to
\begin{equation}    
D_{P,M_k}(o) = \frac{\mathbb{1}\{a = a^\star\}}{\mathrm{g}\left(a^\star \mid w\right)}\left(u(a^\prime, 1,w)  - u(a^\prime, 0,w)\right) \left( m_k - \mathrm{q}(1| a^\star, w) \right). \label{bi_mk}
\end{equation}
According to Theorem \ref{EIF_prime}, the EIF $D_P(o)$ depends on nuisance parameters $\mathrm{b},\mathrm{g},\mathrm{q},\mathrm{r},s,u, \text{and } v$. The EIF has the property of multiple robustness, allowing for consistent estimation even if certain nuisance parameters are inconsistently estimated. We examine the multiple robustness property of the EIF for $\theta_k^\prime$ by deriving the second-order term that quantifies the difference between the true values of nuisance parameters and arbitrary, potentially misspecified values. 

\begin{lemma} (Multiple robustness of EIF for $\theta_k^\prime$) Let $P_1$ represent a probability distribution within the nonparametric model $\mathcal{M}$, which may differ from the true underlying probability distribution $P$. Suppose that any one of the conditions (a)--(e) in Table~\ref{mult} is satisfied, such that the corresponding checked ($\checkmark$) nuisance parameters are consistently estimated. Then, $E(D_{P_1}(o)) = o_P(1)$.

\begin{table}[H]
\captionsetup{justification=raggedright, singlelinecheck=false}
\centering
\renewcommand{\arraystretch}{1.2} % Increases the row height
\begin{tabular}{c c c c c c c c}
\hline
condition & $ \mathrm{b}\left(m_k, a, z, l, w\right) $ & $\mathrm{g}\left(a \mid w\right)$ & $\mathrm{q}(m_k \mid a,w)$ & $\mathrm{r}(m_k \mid a,z,w)$ & $s(a,z,w)$ & $u(a,m_k,w)$ & $v(a,w)$ \\
\hline
(a) & $\checkmark$ & & $\checkmark$ & & $\checkmark$ & & $\checkmark$ \\
(b) & $\checkmark$ & $\checkmark$ & $\checkmark$ & & $\checkmark$ & &  \\
(c) & & & $\checkmark$ & $\checkmark$ & & & $\checkmark$\\
(d) & & $\checkmark$ & $\checkmark$ & $\checkmark$ & & & \\
(e) & $\checkmark$ & $\checkmark$ & & $\checkmark$ & & $\checkmark$ & \\
\hline
\end{tabular}
\caption{Consistency conditions for nuisance parameters. The columns represent the nuisance parameters. Each row specifies a condition where specific combinations of nuisance parameters must be consistently estimated to ensure the consistency of the EIF.} \label{mult}
\end{table}
\end{lemma}
The proof of Lemma 1 is provided in Section 2 of the Supplementary Material. Based on the reparametrisation of Equations (\ref{eqs})--(\ref{eqv}), obtaining consistent estimators for $v(a,w)$, $u(a,m_k,w)$, and $s(a,z,w)$ requires consistent estimators for $\mathrm{b}\left(m_k, a, z, l, w\right)$, $\mathrm{q}(m_k \mid a,w)$, and $\mathrm{r}(m_k \mid a,z,w)$. Consequently, conditions (b) and (e) in Table~\ref{mult} are less interesting because they require the consistent estimation of more nuisance parameters than condition (d). Moreover, condition (a) is equivalent to condition (c), as both require the consistent estimation of $\mathrm{b}(m_k, a, z, l, w)$, $\mathrm{q}(m_k \mid a, w)$, and $\mathrm{r}(m_k \mid a, z, w)$.

Importantly, the inference is not multiply robust \citep{Vansteelandt2007Estimation, Vermeulen2015Bias}. This means that if any nuisance parameters are inconsistently estimated, the standard errors will remain biased despite the property of multiple robustness described above \citep{Vermeulen2015Bias, Dukes2024On}. In the following sections, we derive efficient estimators based on EIF and study their asymptotic properties.

\subsection{One-step estimator for \texorpdfstring{$\theta_k^\prime$}{theta\_k\_prime}}
Let $\hat{\theta}_k^\prime$ denote the plug-in estimator of $\theta_k^\prime$, obtained by substituting estimated components $\hat{P}$ of $P$ into the target estimand (Equation (\ref{t1})). This estimator is generally subject to first-order bias, which can be estimated using the EIF as $-\frac{1}{n}\sum_{i=1}^n D_{\hat{P}}(O_i)$, where $D_{\hat{P}}(O_i)$ represents the EIF estimated using the observed data. Thus the one-step estimator is constructed by subtracting this bias from the plug-in estimator \citep{Whitney2004Twicing}, as follows: ${\hat{\theta}_{k,\text{os}}^\prime} = \hat{\theta}_k^\prime + \frac{1}{n}\sum_{i=1}^n D_{\hat{P}}(O_i)$. According to Equation (\ref{EIF_prime_eq}), ${\hat{\theta}_{k,\text{os}}^\prime}$ can be calculated as:
\begin{equation}
    \hat{\theta}_{k,\text{os}}^\prime = \frac{1}{n}\sum_{i=1}^n \left( D_{\hat{P},Y}(O_i) + D_{\hat{P},Z}(O_i) + D_{\hat{P},M_k}(O_i) + D_{\hat{P},L}(O_i) +  \hat{v}(a^\prime, w_i)\right), \label{one-step}
\end{equation}
where $\hat{v}(a^\prime, w_i)$ denotes the estimate of nuisance parameter $v$ for the $i$-th observation.

An alternative approach to debiasing is the estimating equation estimator, obtained by solving the estimating equation $\frac{1}{n}\sum_{i=1}^n D_{\hat{P}}(O_i)=0$ directly. In our case, since the EIF is linear in $\theta_k^\prime$ according to Equation (\ref{EIF_prime_eq}), deriving the estimating equation is straightforward, resulting in an estimator that is equivalent to the one-step estimator.

The one-step estimator can be implemented by estimating each component of the EIF in Equation (\ref{EIF_prime_eq}), which can be done by fitting the parametric regression or flexible machine learning models to estimate each nuisance parameter. Here we consider the SuperLearner for estimating these nuisance parameters. The SuperLearner is an ensemble learning method that constructs a prediction model by generating a weighted combination of multiple candidate algorithms, with the weights selected to minimise the cross-validated risk function \citep{vanDerLaan2007}. It has been theoretically demonstrated that the SuperLearner asymptotically performs as well as the best possible estimator (oracle selector) given the candidate learners \citep{DUDOIT2005131,vanderLaanDudoitKeles2004}.

When utilising machine learning methods in high-dimensional settings, the one-step estimator ${\theta_{k,\text{os}}^\prime}$ may not achieve asymptotic normality unless so-called Donsker conditions are met \citep{chernozhukov2018double}. To eliminate the need for Donsker conditions, we propose a cross-fitted version of the one-step estimator \citep{zheng2011cross, chernozhukov2018double}. Given the EIF in Equation (\ref{EIF_prime_eq}) is Neyman orthogonal \citep{chernozhukov2018double}, the cross-fitted one-step estimator can be viewed as an implementation of the DML estimator \citep{pfanzagl1982contributions, Díaz2019machine}. 

\subsubsection{Implementation}\label{os-imp}
The cross-fitted one-step estimator can be implemented using the following steps:
\begin{enumerate}
    \item Randomly partition data into $J$ approximately same-sized folds, denoted by $\mathcal{U}_1$, $\mathcal{U}_2$, $\dots$, $\mathcal{U}_J$. For each fold $\mathcal{U}_{\ell}$, let $\mathcal{U}_{\ell}^C$ denote the data from all other folds, i.e. \{$\mathcal{U}_j: j \neq \ell$\}. 
    \item For each fold $\mathcal{U}_{\ell}$, use data in $\mathcal{U}_{\ell}^C$ to train models for the estimation of nuisance parameters. 
    \begin{enumerate}
        \item The models for estimating $\mathrm{b},\mathrm{g},\mathrm{q},\mathrm{r}$ can be directly trained using the SuperLearner. For example, the model for $\mathrm{b}$ can be trained by predicting the outcome $Y$ based on $M_k,A,Z,L, \text{ and } W$. 
        \item To train the model for estimating $s, u, \text{ and } v$, we use the repeated regression method proposed by \cite{Díaz2021nonpara}. For example, to estimate $v(a,w)$, which is reformulated as 
        $$v(a,w) = \mathrm{E} \left( \mathrm{b}\left(m_k, a, z, l, w\right) \frac{\mathrm{q}\left(m_k \mid a^\star, w\right)}{\mathrm{r}(m_k \mid a,z,w)} \mid A=a , W=w\right),$$ we can construct a pseudo-outcome $\mathrm{b}\left(m_k, a, z, l, w\right) \frac{\mathrm{q}\left(m_k \mid a^\star, w\right)}{\mathrm{r}(m_k \mid a,z,w)}$, where $\mathrm{b}$, $\mathrm{q}$, and $\mathrm{r}$ are estimated from step 2(a). Then, regress the pseudo-outcome on $A$ and $W$ to train the model for estimating $v(a, w)$.
    \end{enumerate}
    \item Apply the trained models to the data in $\mathcal{U}_{\ell}$ to obtain estimates of the nuisance functions for each observation. Using these estimates, calculate each component of Equation (\ref{one-step}) for the observations in $\mathcal{U}_{\ell}$.
    \item Average the estimates for components in Equation (\ref{one-step}) over all observations across all folds to obtain the cross-fitted one-step estimator:
     $$
     \hat{\theta}_{k,\text{cf-os}}^\prime = \frac{1}{n}\sum_{j=1}^{J} \sum_{i \in \mathcal{U}_{j}} \left( D_{\hat{P},Y}^{\{j\}}(O_i) + D_{\hat{P},Z}^{\{j\}}(O_i) + D_{\hat{P},M_k}^{\{j\}}(O_i) + D_{\hat{P},L}^{\{j\}}(O_i) +  \hat{v}^{\{j\}}(a^\prime, w_i)\right),
     $$
     where $D_{\hat{P},Y}^{\{j\}}(O_i)$, $D_{\hat{P},Z}^{\{j\}}(O_i)$, $D_{\hat{P},M_k}^{\{j\}}(O_i)$, and $D_{\hat{P},L}^{\{j\}}(O_i)$ represent the estimates of the $Y$, $Z$, $M_k$, and $L$ components of the EIF for observation $i$ in fold $j$ respectively. Additionally, $\hat{v}^{\{j\}}(a^\prime, w_i)$ represents the estimate of the nuisance function $v$ for observation $i$ in fold $j$.
\end{enumerate}

\subsubsection{Asymptotic properties}
Here we characterise the asymptotic properties of the cross-fitted one-step estimator $\hat{\theta}_{k,\text{cf-os}}^\prime$. Define the $L^2$ norm as $\|f\| = (\int |f(o)|^2 \mathrm{d}P(o))^{1/2}$ for a given square-integrable function $f(o)$. We proceed under the following assumptions, which are important for ensuring the regularity conditions required for consistency and asymptotic normality of $\hat{\theta}_{k,\text{cf-os}}^\prime$.

\begin{assumption}(Positivity) Given random variables $W$ and $Z$, there exists a constant $\epsilon > 0$ such that, for each $a', m_k$ and $a^\star,$ the conditional densities $\mathrm{g}(a' \mid W)$, $\mathrm{r}(m_k \mid a', Z, W)$, and $\mathrm{g}(a^\star \mid W)$ are uniformly bounded away from $\epsilon$ with probability 1. 
\end{assumption}

The positivity assumption ensures that these conditional densities are uniformly bounded away from zero. In our LSAC example, these assumptions are plausible, since both $A$ and $M_k$ are binary variables.

\begin{assumption} ($n^{1/2}$ - convergence of second order terms) We assume that:
\begin{equation*}
\begin{aligned}
 & \| \hat{\mathrm{g}} - \mathrm{g} \| \| \hat{v} - v \|   = o_P (n^{-1/2}), \\
 & \| \hat{\mathrm{b}} - \mathrm{b} \| \{ \| \hat{\mathrm{q}} - \mathrm{q} \|  +  \| \hat{\mathrm{r}} - \mathrm{r} \|   \}  = o_P (n^{-1/2}), \\
 & \| \hat{s} - s \|  \| \hat{\mathrm{r}} - \mathrm{r} \|   = o_P (n^{-1/2}),\\
 &  \| \hat{\mathrm{q}} - \mathrm{q} \| \{\| \hat{u} - u \|  + \| \hat{\mathrm{r}} - \mathrm{r} \| + \|\hat{\mathrm{g}} - \mathrm{g} \|  \}  = o_P (n^{-1/2}),
\end{aligned}
\end{equation*}
\end{assumption}
where the terms $\hat{\mathrm{g}}, \hat{v}, \hat{\mathrm{q}}, \hat{\mathrm{b}}, \hat{\mathrm{r}}, \hat{s}, \hat{u}$ denote corresponding estimators of the true parameters $\mathrm{g}, v, \mathrm{q}, \mathrm{b}, \mathrm{r}, s, u$.
This assumption is satisfied if the nuisance parameters converge to their true values at a rate faster than $n^{-1/4}$ (but can also hold under weaker conditions). This rate is slower than the parametric rate and can be achieved by flexible machine learning methods, such as Lasso \citep{Bickel2009, Alexandre2013Least}, highly-adaptive Lasso \citep{Benkeser2016}, and a class of regression trees and random forests \citep{wager2016adaptiveconcentrationregressiontrees} in high-dimensional settings, provided the true nuisance parameters have sufficient smoothness. 

\begin{theorem}
    Under Assumptions 1 and 2, we have 
    $$ \sqrt{n} (\hat{\theta}_{k,\text{cf-os}}^\prime - \theta_k^\prime) \to N\left(0, \mathrm{var}(D_P(O))\right).$$\label{norm_prime}
\end{theorem}
Of note, $\mathrm{var}(D_P(O))$ is the nonparametric efficiency bound. Thus, this theorem shows that under the above assumptions, the cross-fitted one-step estimator is root-n consistent, and $\sqrt{n} (\hat{\theta}_{k,\text{cf-os}}^\prime - \theta_k^\prime)$ asymptotically follows a zero-mean normal distribution with a variance that attains the nonparametric efficiency bound. The proof of Theorem \ref{norm_prime} is provided in Section 3 of the Supplementary Material. According to Theorem \ref{norm_prime}, the standard error can be estimated using the sample variance of the EIF, and this can be used to obtain Wald-type confidence intervals. 

In practice, even if the positivity assumption is satisfied, we may still observe large inverse probability weights when estimating the one-step estimator due to small estimated conditional probabilities for nuisance parameters $\mathrm{g}$ and $\mathrm{r}$. These large weights can lead to high variance in the estimated EIF, which will reduce the efficiency and lead to unstable estimates. To mitigate the issue, the weights are stabilised by dividing each component of the EIF in Equation (\ref{EIF_prime_eq}) by the corresponding empirical mean of the weights \citep{Díaz2021nonpara}. For example, for the component $D_{P,Y}(o)$, weight stabilisation is applied by dividing $D_{P,Y}(o)$ by the empirical mean of $\mathbb{1}\{a = a^\prime\}/\mathrm{g}\left(a^\prime \mid w\right)$ $\times \mathrm{q}\left(m_k \mid a^\star, w\right) / \mathrm{r}(m_k \mid a^\prime, z, w)$. Since the sample average of the weights asymptotically converges to 1, both Theorem~\ref{EIF-theory} and the alternative representation of the EIF defined in Section~\ref{alternative-eif} remain valid despite this weight stabilisation.

\subsection{Partial TML estimator for \texorpdfstring{$\theta_k^\prime$}{theta\_k\_prime}} \label{pTMLE}
An alternative to the one-step estimator is the TML estimator, which constructs the efficient estimator by tuning nuisance parameters to ensure that the empirical means of the components $D_{P,Y}(o)$, $ D_{P,Z}(o)$, $D_{P,M_k}(o)$, and $D_{P,L}(o)$ in the EIF are set to zero \citep{van2006targeted, van2011targeted}. However, constructing TML estimators is challenging for $\hat{\theta}_k^\prime$ as $D_{P,L}(o)$ and $D_{P,Z}(o)$ involve high-dimensional conditional densities for $L$ and $Z$ respectively, as in the LSAC example. 

Building on the concept of the partial TML estimator introduced by \cite{Rudolph2024Practical}, we develop a partial TML estimator that only targets the components $D_{P,Y}(o)$ and $D_{P,M_k}(o)$ of the EIF, ensuring their empirical means are equal to zero, with a focus on the case of binary $M_k$. Our partial TML procedure for $D_{P,M_k}(o)$ is derived based on the simplification of $D_{P,M_k}(o)$ when $M_k$ is binary, as described in Equation (\ref{bi_mk}). 

The remaining components of the EIF are calculated the same as the one-step estimator. The partial TML estimator is detailed in Section 4 of the Supplementary Material.

The partial TML estimator is expected to have enhanced finite sample performance relative to the one-step estimator, especially when the estimated conditional probabilities specified in Assumption 1 are small, which can lead to extreme or unstable weights \citep{petersen2012diagnosing, Rudolph2024Practical}. Of note, it is important to include the $Y$ component $D_{P,Y}(o)$ in the partial TML estimator as it is expected to be the predominant source of the overall variability of the estimator due to the weights \citep{Rudolph2024Practical}. 

\subsection{Efficient estimation for \texorpdfstring{$\theta_k$}{theta\_k} and \texorpdfstring{$\theta_{all}$}{theta\_{all}}} 

In this section, we describe the EIF and efficient estimators for the estimands $\theta_k$ and $\theta_{all}$, which can be obtained from the results for $\theta_k^\prime$ by considering special simpler cases. Specifically, the identification formula for $\theta_k^\prime$ reduces to that of $\theta_k$ if $L$ is empty and $Z$ represents all mediators in $M$ except for $M_k$. Further, the identification formula for $\theta_k$ reduces to that of $\theta_{all}$ by replacing $M_k$ with $M$ (given the hypothetical intervention defining $\theta_{all}$ shifts the joint distribution of all mediators) and $Z$ is empty. By leveraging these simplifications, the efficient influence functions for $\theta_k$ and $\theta_{all}$ can be directly obtained from the results for $\theta_k^\prime$. Detailed results for the EIF and efficient estimators for $\theta_k$ and $\theta_{all}$ are provided in Section 5 of the Supplementary Material. Unlike $\theta_k^\prime$, a (full) TML estimator can be derived for both $\theta_k$ and $\theta_{all}$.

\section{Application to the LSAC Study}
In this section, we describe our application of the developed one-step and (partial) TML estimators to the LSAC study. As described in Section \ref{mot-exa}, we drew data on $n=978$ adolescents from LSAC wave 6, wave 7, and its interpolated CheckPoint study. Our overall goal was to assess the extent to which mediator interventions shifting the distribution of non-inflammatory adverse metabolomic markers and high inflammatory burden could mitigate the adverse causal effect of overweight or obesity on high blood pressure in adolescence. We considered the three hypothetical interventions described in Section \ref{med-int} to address the specific research questions (i) and (ii) described in Section~\ref{mot-exa}. The corresponding interventional indirect effects defined in Section \ref{tar-est} were estimated. When estimating $\theta_k^\prime$, we established a causal ordering for the metabolites by determining whether they were likely causal ancestors or descendants of GlycA following \cite{Feingold2022}. A list of ordered metabolites is provided in Section 6 of the Supplementary Material.

To implement our proposed estimators, we extended the R package \textit{medoutcon} \citep{Hejazi2022}, developing an R package called \textit{medoutconRCT}. We used the R package \textit{sl3} \citep{coyle2021sl3-rpkg} to implement a SuperLearner ensemble that combined a generalised linear model, Lasso \citep{Tibshirani1996}, elastic net \citep{Friedman2010}, a single hidden layer neural network \citep{Ripley1996}, random forest \citep{Wright2017}, highly adaptive Lasso \citep{Benkeser2016}, and XGBoost \citep{Chen2016}. We applied both 5-fold and 10-fold cross-fitting to estimate the three interventional indirect effects. Since the outcome is rare (5\% prevalence), we used stratified cross-fitting, with the folds stratified by the outcome. Given random seeds could also substantially influence machine learning-based causal effect estimation, we followed \cite{Schader2024Don} and averaged the estimates over $10$ replicates using different seeds. Extreme estimates of IIE were excluded because they resulted from non-convergence. Of note, our ensemble was computationally intensive, requiring approximately 10 hours of computation on a single CPU core with 16GB of memory to estimate $\theta_k^\prime$ using a 10-fold cross-fitting procedure.

The results are presented in Figure \ref{fig-re}, with detailed point estimates and confidence intervals provided in Table 2 of the Supplementary Material. The total causal effect, calculated as a risk difference, was 0.12 (95\% CI: 0.06, 0.19). That is, we estimated that there were 12 (95\% CI: 6-19) additional cases of high blood pressure per 100 when adolescents were exposed to overweight or obesity compared to when they were not.

A hypothetical intervention shifting the joint distribution of all mediators in adolescents with overweight or obesity (exposed) to that without (unexposed) produced the largest estimates of interventional indirect effect, suggesting that such an intervention could reduce the heightened risk of high blood pressure under exposure by about 7 cases per 100 adolescents, with a 95\% confidence interval ranging from about 1–2 to 12–13 cases per 100 adolescents, depending on the GlycA cutoff and the estimation method used. In contrast, a hypothetical intervention shifting only the distribution of high inflammatory burden among those exposed to the level in the unexposed, without accounting for flow-on effects, resulted in the smallest estimates of interventional indirect effect, indicating that such an intervention could reduce the risk of high blood pressure by 2-4 cases per 100 adolescents, with a 95\% confidence interval ranging from about -3–0 to 6–8 cases per 100 adolescents, when considering the 50th percentile cutoff for GlycA, with no observable risk reduction for the 75th percentile cutoff. However, when accounting for flow-on effects, estimates suggest that such a hypothetical intervention could achieve a risk reduction only slightly smaller than the intervention shifting the joint distribution of all mediators, ranging from 5-7 cases per 100 adolescents, with a 95\% confidence interval ranging from about -2–3 to 10–13 cases per 100 adolescents, depending on the GlycA cutoff and the estimation method used. This highlights the importance of accounting for flow-on effects when mediators are correlated, although this depends on a causal ordering assumption that may not always be straightforward to make.

In this example, the one-step and (partial) TML estimator provided similar estimates across all scenarios, and additionally, increasing the number of cross-fitting folds from 5 to 10 also had little impact on the estimates, in line with recent simulation evidence in simpler settings \citep{meng2022refine2toolevaluaterealworld, ellul2024causal}. Per above, the choice of GlycA cutoff used to define high inflammatory burden (50th or 75th percentile) did not substantially change findings for two of the three estimands considered. However, using the 75th percentile cutoff resulted in larger variances, likely due to limited sample size in certain strata.

The above results suggest the need for future research and intervention development aiming at reducing heightened inflammation and other adverse metabolic profiles, as they could play a substantial role in reducing the adverse effects of overweight or obesity on later cardiovascular disease risk. Nonetheless, in this as well as other applications of this methodology, it is important to interpret findings alongside a consideration of the relevance of each of the estimands given the context.

A hypothetical intervention targeting the joint distribution of all mediators is likely to offer the greatest potential impact (when all mediators act on the outcome in the same direction). The corresponding estimand $\text{IIE}_{all}$ is most relevant when we can conceive of an intervention capable of modifying all mediators simultaneously. This is unlikely to be the case in the LSAC study, where an intervention jointly changing all mediators is challenging to imagine. 

The estimand $\text{IIE}_k$ is most relevant when the interrelationships among mediators are expected to be weak. In such cases, shifting the distribution of each mediator is not expected to have important flow-on effects on causally descendant mediators. This assumption is not plausible for studies like our LSAC study, where non-inflammatory adverse metabolomic markers and the inflammatory marker GlycA are moderately correlated. In this context, intervening on GlycA is likely to have a flow-on effect on other metabolomic markers. However, the required causal assumptions for this estimand are weaker, as no causal ordering needs to be specified. This is an important consideration, especially when the interplay among mediators is not well understood, which is likely to be the case in high-dimensional biological studies. 

The estimand $\text{IIE}_k^\prime$ is most relevant when mediators are correlated and a clear causal ordering among them can be established. Accounting for flow-on effects offers a more realistic understanding of the impact of a hypothetical intervention. Although it may be challenging to establish causal ordering in high-dimensional mediation settings, in cases like our LSAC study where a specific mediator is of interest, we just need to determine whether a mediator is a causal ancestor or descendant of the mediator of interest. This may be a more feasible endeavour than establishing a complete order for all the mediators.

\begin{figure}[ht]
    \centering
    \includegraphics[width=\textwidth]{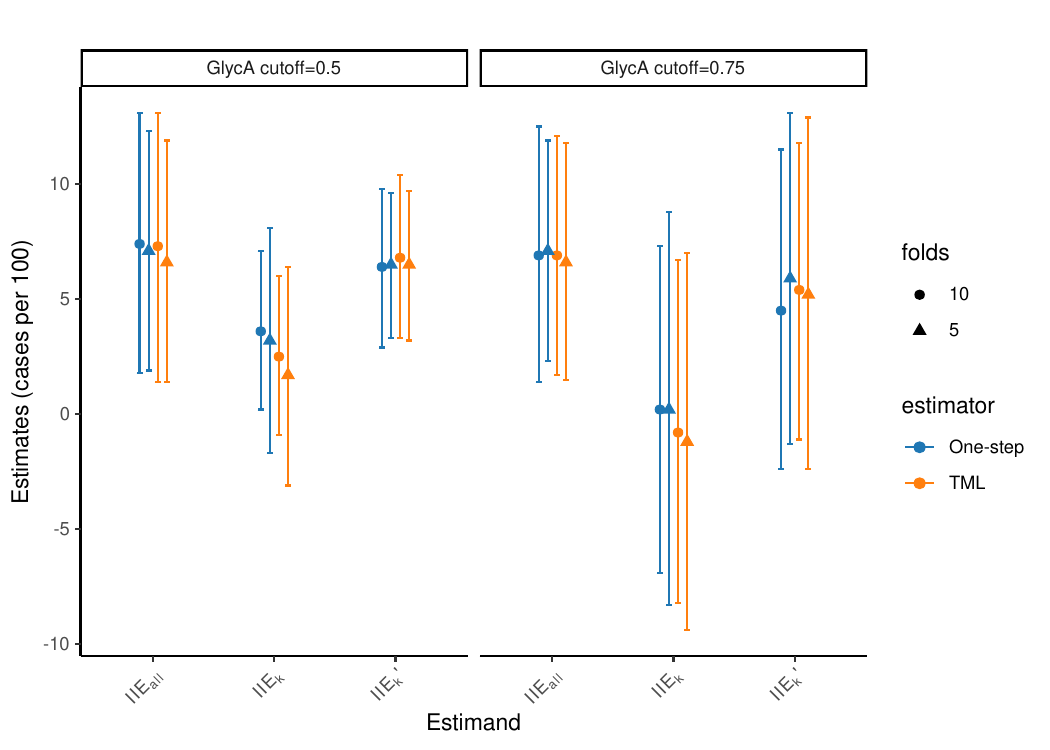}
    \caption{Estimated interventional indirect effects for different estimands ($\text{IIE}_{all}$, $\text{IIE}_k$, and $\text{IIE}_k^\prime$), presented along with 95\% confidence intervals for different cross-fitting folds (5 and 10) and GlycA cutoff levels (0.5 and 0.75). For $\text{IIE}_k^\prime$, TML refers to the partial TML estimator as proposed in Section~\ref{pTMLE}.} 
    \label{fig-re}
\end{figure}

\section{Discussion}
We developed efficient estimation methods, based on one-step and (partial) TML estimators, for interventional effects mapped to a target trial. These estimators are referred to as causal machine learning estimators when combined with machine learning techniques, which we leverage to tackle high-dimensional mediator settings. This work extends prior work examining causal machine learning methods in causal mediation analysis \citep{Díaz2021nonpara, Benkeser2021nonpara, Rudolph2024Practical, ran2024nonparametricmotioncontrolfunctional} to the estimation of target estimands that are key for epidemiological studies: interventional mediation effects defined explicitly in terms of hypothetical mediator interventions of interest \citep{Moreno-Betancur2021mediation, Dashti2022, Afshar2024}. Our approach advances current methods for the estimation of these effects by addressing the challenges of high-dimensional mediation.

Specifically, the interventional effects we consider represent the impact of hypothetical mediator interventions that shift the distribution of mediators both individually and jointly. Given the potential correlation between mediators, an intervention shifting the distribution of a mediator individually may have flow-on effects on its causal descendants. We therefore develop efficient estimators for interventional effects that account for these flow-on effects when a specific causal ordering is known or can be assumed. We also consider interventional effects representing the impact of shifting a mediator individually, without accounting for flow-on effects (which may be necessary when a causal ordering is unknown). In this case, the target estimand is equivalent to that described by \cite{Díaz2021nonpara} in the setting with high-dimensional intermediate confounders and low-dimensional mediators. When a hypothetical intervention shifts the joint distribution of all mediators, the target estimand is similar to those considered by \cite{Tchetgen2012Semiparametric} and \cite{Farbmacher2022Causal}.

Our proposed estimators and implementations for the three target estimands are applicable to settings where the exposure $A$ is binary and the outcome variable $Y$ is either binary or continuous. For $\theta_k^\prime$ and $\theta_k$, the mediator of interest $M_k$ needs to be binary, but the remaining mediators ($M_1, \dots, M_{k-1}, M_{k+1}, \dots, M_K$) can be of any type. In contrast, for $\theta_{all}$, there are no restrictions on mediator types. Future work is needed to extend and implement these estimators to other settings. Here we outline some possible considerations for these potential extensions. If $M_k$ is a continuous variable, the corresponding conditional densities could be estimated using parametric or machine learning methods as suggested by \cite{Díaz2021nonpara}. Additionally, if $M_k$ is a multivariate vector, the reparameterisation and implementation proposed by \cite{Rudolph2024Practical} could be considered. Furthermore, it may also be possible to directly estimate each component, such as $D_{P,Y}(o)$, in the EIF using automatic debiased machine learning \citep{Chernozhukov2022debiasedML, liu2024generaltargetedmachinelearning}. However, this requires the specification of custom loss functions in machine learning methods which is only supported in limited software \citep{liu2024generaltargetedmachinelearning}.

Although we have illustrated the application of the proposed methods to a real-world longitudinal cohort study with high-dimensional mediators, challenges in their practical implementation remain, including providing guidance on selecting machine learning algorithms and their tuning parameters in SuperLearner, as well as handling missing data, see (e.g., \cite{DashtiHandling2024, LevisRobust}). More work is needed to provide practical guidance on optimising the estimation of these practically relevant interventional effects by evaluating the proposed estimators in realistic simulation studies. We plan to address these gaps in future work.

There are an increasing number of novel methods for causal mediation analysis, but many of these methods do not estimate parameters that are directly relevant to informing decision-making in clinical medicine and population health. Our study improves the practical utility of causal machine learning methods for mediation analysis by proposing causal machine learning methods for estimating practically relevant target estimands in the high-dimensional mediation context and thus bridging the gap between methodological advances and their potential to inform decision-making and clinical interventions.

%\section*{ACKNOWLEDGMENTS}
\section*{ACKNOWLEDGMENTS} 

This work was supported by an Investigator Grant fellowship to MMB [grant ID: 2009572] and to DB [grant ID: 1175744] from the National Health and Medical Research Council. The Murdoch Children’s Research Institute is supported by the Victorian Government's Operational Infrastructure Support Program.

\bibliographystyle{abbrvnat}
\bibliography{references} 
\clearpage

\begin{titlepage}
   \begin{center}
       \vspace*{1cm}

       \Huge{SUPPLEMENTARY MATERIAL\\}
       \vspace{3cm}
       \LARGE{
       Causal machine learning for high-dimensional mediation analysis using interventional effects mapped to a target trial}\\
       \vspace{4cm}
       \Large{Tong Chen$^{1,2}$, Stijn Vansteelandt$^{3}$, David Burgner$^{4,5}$, 
       Toby Mansell$^{4,5}$, \\ Margarita Moreno-Betancur$^{2,4}$}\\
       \vspace{3cm}

       \large{1. Melbourne Dental School, University of Melbourne\\ 
       2. Clinical Epidemiology and Biostatistics Unit, Murdoch Children’s Research Institute\\
       3. Department of Applied Mathematics, Computer Science and Statistics, Ghent University\\
       4. Department of Paediatrics, University of Melbourne\\
       5. Inflammatory Origins, Murdoch Children’s Research Institute}

       \vspace{3cm}
       
       \Large{\today}
            
   \end{center}
\end{titlepage}

\setcounter{section}{0}
\setcounter{table}{0}
\section{Efficient influence function for \texorpdfstring{$\theta_k^\prime$}{theta\_k\_prime} (proof of Theorem 4.1)}

\begin{proof}
We derive the efficient influence function using a parametric submodel, defined by perturbations parameterised through a one-dimensional mixture model \citep{hines2022demystifying}. Let $\Psi(P_t)$ denote the parametric submodel for $\theta_k^\prime$. Under the parametric submodel, we have
\begin{equation*}
\begin{aligned} &
\Psi(\mathcal{P}_t) = \int \mathrm{b}_t(m_k, a^{\prime}, z, l, w)  \times \mathrm{p}_t(z|a^\prime,w)  \times \mathrm{q}_t(m_k|a^\star,w)  \times \mathrm{p}_t(l|a^\prime,w,z, m_k) \times \mathrm{p}_t(w) \mathrm{d}m_k \mathrm{d}z \mathrm{d}l \mathrm{d}w .
\end{aligned}
\end{equation*}

Applying the derivative operator gives
\begin{equation*}
\begin{aligned} &
\partial \Psi(\mathcal{P}_t) = \int \partial_t \mathrm{b}_t(m_k, a^{\prime}, z, l, w)  \times \mathrm{p}_t(z|a^\prime,w)  \times \mathrm{q}_t(m_k|a^\star,w)  \times \mathrm{p}_t(l|a^\prime,w,z, m_k) \times \mathrm{p}_t(w) \\ & +
 \mathrm{b}_t(m_k, a^{\prime}, z, l, w)  \times \partial_t \mathrm{p}_t(z|a^\prime,w)  \times \mathrm{q}_t(m_k|a^\star,w)  \times \mathrm{p}_t(l|a^\prime,w,z, m_k) \times \mathrm{p}_t(w) \\ & +
\mathrm{b}_t(m_k, a^{\prime}, z, l, w)  \times  \mathrm{p}_t(z|a^\prime,w)  \times \partial_t \mathrm{q}_t(m_k|a^\star,w)  \times \mathrm{p}_t(l|a^\prime,w,z, m_k) \times \mathrm{p}_t(w) \\ & +
\mathrm{b}_t(m_k, a^{\prime}, z, l, w)  \times  \mathrm{p}_t(z|a^\prime,w)  \times \mathrm{q}_t(m_k|a^\star,w)  \times \partial_t \mathrm{p}_t(l|a^\prime,w,z, m_k) \times \mathrm{p}_t(w) \\ & +
\mathrm{b}_t(m_k, a^{\prime}, z, l, w)  \times  \mathrm{p}_t(z|a^\prime,w)  \times \mathrm{q}_t(m_k|a^\star,w)  \times \mathrm{p}_t(l|a^\prime,w,z, m_k) \times \partial_t \mathrm{p}_t(w)
\mathrm{d}m_k \mathrm{d}z \mathrm{d}l \mathrm{d}w .
\end{aligned}
\end{equation*}

Evaluating these derivative operators
\begin{equation*}
\begin{aligned} &
\partial \Psi(\mathcal{P}_t) = \int \Bigg[ \frac{\mathbb{1}\{ (m_k, a^\prime, z, l, w)= \tilde{o}\}}{\mathrm{p}\left(m_k, a^\prime, z, l, w\right)} \left(y - \mathrm{b}(m_k, a^{\prime}, z, l, w)\right)    \times \mathrm{p}_t(z|a^\prime,w)  \times \mathrm{q}_t(m_k|a^\star,w)  \times \mathrm{p}_t(l|a^\prime,w,z, m_k) \times \mathrm{p}_t(w) \\ 
& + \mathrm{b}_t(m_k, a^{\prime}, z, l, w)  \times \frac{\mathbb{1}\{ (a^\prime, w) = \tilde{o}\}}{\mathrm{p}\left(a^\prime, w\right)} \left({\mathbb{1}\{z = \tilde{z} \}-\mathrm{p}\left(z \mid a^\prime, w\right)} \right)  \times \mathrm{q}_t(m_k|a^\star,w)  \times \mathrm{p}_t(l|a^\prime,w,z, m_k) \times \mathrm{p}_t(w) \\ 
& + \mathrm{b}_t(m_k, a^{\prime}, z, l, w)  \times  \mathrm{p}_t(z|a^\prime,w)  \times \frac{\mathbb{1}\{(a^\star, w)= \tilde{o}\}}{\mathrm{p}\left(a^\star, w\right)} \left({\mathbb{1}\{m_k = \tilde{m_k}\}-\mathrm{q}\left(m_k \mid a^\star, w\right)} \right)  \times \mathrm{p}_t(l|a^\prime,w,z, m_k) \times \mathrm{p}_t(w) \\ & +
\mathrm{b}_t(m_k, a^{\prime}, z, l, w)  \times  \mathrm{p}_t(z|a^\prime,w)  \times \mathrm{q}_t(m_k|a^\star,w)  \times \frac{\mathbb{1}\{(m_k, a^\prime, z, w) = \tilde{o}\}}{\mathrm{p}\left(m_k, a^\prime, z, w\right)} \left({\mathbb{1}\{l = \tilde{l}\}-\mathrm{p}\left(l \mid a^\prime, w, z, m_k\right)}\right) \times \mathrm{p}_t(w) \\ & +
\mathrm{b}_t(m_k, a^{\prime}, z, l, w)  \times  \mathrm{p}_t(z|a^\prime,w)  \times \mathrm{q}_t(m_k|a^\star,w)  \times \mathrm{p}_t(l|a^\prime,w,z, m_k) \times \left(\mathbb{1}\{w = \tilde{w}\}-\mathrm{p}(w) \right) \Bigg]
\mathrm{d}m_k \mathrm{d}z \mathrm{d}l \mathrm{d}w. 
\end{aligned}
\end{equation*}

Evaluating the integral results in the EIF
\begin{equation*}
\begin{aligned}&
\frac{\mathbb{1}\{a = a^\prime\}}{\mathrm{g}\left(a^\prime \mid w\right) } \frac{\mathrm{q}\left(m_k \mid a^\star, w\right)}{\mathrm{r}(m_k \mid a^\prime, z, w)} \left( y - \mathrm{b}(m_k, a^\prime, z, l, w) \right)  \\&
+ \frac{\mathbb{1}\{a = a^\prime\}}{\mathrm{g}\left(a^\prime \mid w\right) } \left(  \int \mathrm{b}(m_k, a^\prime, z, l, w) \mathrm{q}(m_k\mid a^\star, w) \mathrm{p}(l\mid a^\prime,w,z,m_k) \mathrm{d}m_k\mathrm{d}l  - v(a^\prime,w) \right)  \\&
+ \frac{\mathbb{1}\{a = a^\star\}}{\mathrm{g}\left(a^\star \mid w\right)}\left(u(a^\prime,m_k,w)  -  \int u(a^\prime,m_k,w) \mathrm{q}(m_k \mid a^\star, w) \mathrm{d}m_k \right)  \\&
+ \frac{\mathbb{1}\{a = a^\prime\}}{\mathrm{g}\left(a^\prime \mid w\right) } \frac{\mathrm{q}\left(m_k \mid a^\star, w\right)}{\mathrm{r}(m_k \mid a^\prime, z, w)} \left(\mathrm{b}(m_k, a^\prime, z, l, w) - \int \mathrm{b}(m_k, a^\prime, z, l, w) \mathrm{p}(l \mid a^\prime,w,z,m_k) \mathrm{d}l\right)  \\&
+ v(a^\prime,w) - \theta_k^\prime,
\end{aligned}
\end{equation*}
where 
\begin{equation*}
\begin{aligned}
s(a,z,w) &=  \int \mathrm{b}(m_k, a, z, l, w) \mathrm{q}(m_k\mid a^\star, w) \mathrm{p}(l\mid a,w,z,m_k)  \mathrm{d}m_k\mathrm{d}l\\
u(a, z, w) &=\int \mathrm{b}\left(m_k, a, z, l, w\right) f\left(z \mid a, w\right) f\left(l \mid a, w, z, m_k\right) \mathrm{d} z \mathrm{d} l, \\
v(a, w) &=\int \mathrm{b}\left(m_k, a, z, l, w\right) f(z \mid a, w) f\left(l \mid a, w, z, m_k\right) f\left(m_k \mid a^\star, w\right) \mathrm{d} m_k \mathrm{d} z \mathrm{d} l.
\end{aligned}
\end{equation*}
\end{proof}

\section{Second-order term for \texorpdfstring{$\theta_k^\prime$}{theta\_k\_prime}}

\begin{proof}
For notational simplicity, the dependence of all functions on $w$ is omitted. Let $P_1 \in \mathcal{M}$ and $\mathrm{p}_w$ be the distribution of $W$. For fixed $a^\star$ and $a^\prime$, we have
\begin{align*} 
& E (D_{P_1} (o)) = \\
& \int \frac{\mathrm{g}(a^\prime)}{\mathrm{g}_1\left(a^\prime\right) } \frac{\mathrm{q}_1\left(m_k \mid a^\star\right)}{\mathrm{r}_1(m_k \mid a^\prime, z)} \left(\mathrm{b}(m_k, a^\prime, z, l) - \mathrm{b}_1(m_k, a^\prime, z, l) \right) \mathrm{p}(l \mid z,a^\prime,m_k) \mathrm{r}(m_k \mid  z,a^\prime) \mathrm{p}(z \mid a^\prime) \mathrm{d}\mathrm{P}_W \mathrm{d}m_k\mathrm{d}z\mathrm{d}l \tag{2.1}  \\+&
\int \frac{\mathrm{g}(a^\prime)}{\mathrm{g}_1\left(a^\prime\right)} s_1(a,z)  \mathrm{p}(z\mid a^\prime) \mathrm{d}\mathrm{P}_W \mathrm{d}m_k\mathrm{d}l\mathrm{d}z  \tag{2.2} \\&
- \int \frac{\mathrm{g}(a^\prime)}{\mathrm{g}_1\left(a^\prime\right)}   v_1(a^\prime) \mathrm{d}\mathrm{P}_W \mathrm{d}m_k\mathrm{d}l  \mathrm{d}z \tag{2.3}  \\+&
\int \frac{\mathrm{g}(a^\star)}{\mathrm{g}_1\left(a^\star\right)} u_1(a^\prime,m_k) \left(\mathrm{q}(m_k \mid a^\star)- \mathrm{q}_1(m_k \mid a^\star)\right)\mathrm{d}\mathrm{P}_W \mathrm{d}z\mathrm{d}l \mathrm{d}m_k    \tag{2.4}  \\+&
\int \frac{\mathrm{g}(a^\prime)}{\mathrm{g}_1(a^\prime)} \frac{\mathrm{q}_1(m_k \mid a^\star)}{\mathrm{r}_1(m_k \mid a^\prime, z)} \left( \mathrm{b}_1(m_k, a^\prime, z, l)\mathrm{p}(l \mid a^\prime,z,m) - \mathrm{b}_1(m_k, a^\prime, z, l)\mathrm{p_1}(l \mid a^\prime,z,m)\right)\mathrm{r}(m_k \mid z, a^\prime) \mathrm{p}(z \mid a^\prime) \mathrm{d}\mathrm{P}_W  \mathrm{d}m_k\mathrm{d}z\mathrm{d}l \tag{2.5} \\+&
\int v_1(a^\prime) \mathrm{d}\mathrm{P}_W \mathrm{d}l\mathrm{d}m_k \mathrm{d}z - 
\int v(a^\prime) \mathrm{d}\mathrm{P}_W \mathrm{d}m_k\mathrm{d}z\mathrm{d}l. \tag{2.6}\\
\end{align*}

We start by noting that

\begin{align*}
(2.3) + (2.6) & = \int \left( v_1\left(a^\prime\right) - v\left(a^\prime\right) \right) \left(1 - \frac{\mathrm{g}\left(a^\prime\right)}{\mathrm{g}_1\left(a^\prime\right)}\right) \mathrm{d}\mathrm{P}_W \mathrm{d}m_k \mathrm{d}z \mathrm{d}l \tag{2.7} \\
& - \int v\left(a^\prime\right) \frac{\mathrm{g}\left(a^\prime\right)}{\mathrm{g}_1\left(a^\prime\right)} \mathrm{d}\mathrm{P}_W \mathrm{d}m_k \mathrm{d}z \mathrm{d}l.
\end{align*}

By expanding the last term, we have
\begin{align*}
&\int v\left(a^\prime\right) \frac{\mathrm{g}\left(a^\prime\right)}{\mathrm{g}_1\left(a^\prime\right)} \mathrm{d}\mathrm{P}_W \mathrm{d}m_k \mathrm{d}z \mathrm{d}l \\
&= \int \frac{\mathrm{g}\left(a^\prime\right)}{\mathrm{g}_1\left(a^\prime\right)} \frac{\mathrm{q}(m_k \mid a^\star)}{\mathrm{r}(m_k \mid a^\prime, z)} \left( \mathrm{b}(m_k, a^\prime, z, l) - \mathrm{b}_1(m_k, a^\prime, z, l) \right) \mathrm{p}(z \mid a^\prime) \mathrm{p}(l \mid a^\prime, z, m_k) \mathrm{r}(m_k \mid a^\prime, z) \mathrm{d}\mathrm{P}_W \mathrm{d}m_k \mathrm{d}z \mathrm{d}l \\
& + \int \frac{\mathrm{g}\left(a^\prime\right)}{\mathrm{g}_1\left(a^\prime\right)} \frac{\mathrm{q}(m_k \mid a^\star)}{\mathrm{r}(m_k \mid a^\prime, z)} \mathrm{b}_1(m_k, a^\prime, z, l) \mathrm{p}(z \mid a^\prime) \mathrm{p}(l \mid a^\prime, z, m_k) \mathrm{r}(m_k \mid a^\prime, z) \mathrm{d}\mathrm{P}_W \mathrm{d}m_k \mathrm{d}z \mathrm{d}l. \tag{2.8}
\end{align*}

As a result, we have
{\small
\begin{align*}
&(2.1) + (2.3) + (2.6)  = (2.7) - (2.8)  \\+
& \int \frac{\mathrm{g}\left(a^\prime\right)}{\mathrm{g}_1\left(a^\prime\right)} \left( \frac{\mathrm{q}_1(m_k \mid a^\star)}{\mathrm{r}_1(m_k \mid a^\prime, z)} - \frac{\mathrm{q}(m_k \mid a^\star)}{\mathrm{r}(m_k \mid a^\prime, z)} \right) \left( \mathrm{b}(m_k, a^\prime, z, l) - \mathrm{b}_1(m_k, a^\prime, z, l) \right) \mathrm{p}(z \mid a^\prime) \mathrm{p}(l \mid a^\prime, z, m_k) \mathrm{r}(m_k \mid a^\prime, z) \mathrm{d}\mathrm{P}_W \mathrm{d}m_k \mathrm{d}z \mathrm{d}l.
 \end{align*}
}

We then calculate 

\begin{align*}
(2.2) - (2.8) &= \int \frac{\mathrm{g}\left(a^\prime\right)}{\mathrm{g}_1\left(a^\prime\right)} \mathrm{p}(z \mid a^\prime) \left( s_1(a^\prime, z) - s(a^\prime, z) \right) \mathrm{d}\mathrm{P}_W \mathrm{d}m_k \mathrm{d}z \mathrm{d}l +\\
&\int \frac{\mathrm{g}\left(a^\prime\right)}{\mathrm{g}_1\left(a^\prime\right)}  \left( \mathrm{b}(m_k, a^\prime, z, l) - \mathrm{b}_1(m_k, a^\prime, z, l) \right) \mathrm{q}(m_k \mid a^\star) \mathrm{p}(z \mid a^\prime) \mathrm{p}(l \mid a^\prime, z, m_k) \mathrm{d}\mathrm{P}_W \mathrm{d}m_k \mathrm{d}z \mathrm{d}l
\end{align*}

We can then obtain 

\small{
\begin{align*}
&(2.2) + (2.5) - (2.8) = \int \frac{\mathrm{g}\left(a^\prime\right)}{\mathrm{g}_1\left(a^\prime\right)} (s(a^\prime, z) - s_1(a^\prime, z)) \left( \frac{\mathrm{r}(m_k \mid a^\prime, z)}{\mathrm{r}_1(m_k \mid a^\prime, z)} - 1\right) \mathrm{p}(z \mid a^\prime)  \mathrm{d}\mathrm{P}_W \mathrm{d}m_k \mathrm{d}z \mathrm{d}l \tag{2.9}\\
& + \int \frac{\mathrm{g}\left(a^\prime\right)}{\mathrm{g}_1\left(a^\prime\right)}  \left( \mathrm{b}(m_k, a^\prime, z, l) - \mathrm{b}_1(m_k, a^\prime, z, l) \right) \mathrm{q}(m_k \mid a^\star) \mathrm{p}(z \mid a^\prime) \mathrm{p}(l \mid a^\prime, z, m_k) \mathrm{d}\mathrm{P}_W \mathrm{d}m_k \mathrm{d}z \mathrm{d}l \\
& + \int \frac{\mathrm{g}\left(a^\prime\right)}{\mathrm{g}_1\left(a^\prime\right)}  \frac{\mathrm{r}(m_k \mid a^\prime, z)}{\mathrm{r}_1(m_k \mid a^\prime, z)} \left( \mathrm{b}_1(m_k, a^\prime, z, l)\mathrm{q}_1(m_k \mid a^\star) - \mathrm{b}(m_k, a^\prime, z, l) \mathrm{q}(m_k \mid a^\star)\right)  \mathrm{p}(z \mid a^\prime) \mathrm{p}(l \mid a^\prime, z, m_k) \mathrm{d}\mathrm{P}_W \mathrm{d}m_k \mathrm{d}z \mathrm{d}l \\
& = (2.9) + \int \frac{\mathrm{g}\left(a^\prime\right)}{\mathrm{g}_1\left(a^\prime\right)}  \left( \mathrm{b}_1(m_k, a^\prime, z, l) - \mathrm{b}(m_k, a^\prime, z, l) \right)  \left(  \frac{\mathrm{r}(m_k \mid a^\prime, z)}{\mathrm{r}_1(m_k \mid a^\prime, z)} \mathrm{q}_1(m_k \mid a^\star) - \mathrm{q}(m_k \mid a^\star) \right)  \mathrm{p}(z \mid a^\prime) \mathrm{p}(l \mid a^\prime, z, m_k) \mathrm{d}\mathrm{P}_W \mathrm{d}m_k \mathrm{d}z \mathrm{d}l \tag{2.10}\\
& + \int \frac{\mathrm{g}\left(a^\prime\right)}{\mathrm{g}_1\left(a^\prime\right)} \frac{\mathrm{r}(m_k \mid a^\prime, z)}{\mathrm{r}_1(m_k \mid a^\prime, z)}  \left( \mathrm{q}_1(m_k \mid a^\star)  -  \mathrm{q}(m_k \mid a^\star) \right)  u(a,m_k) \mathrm{d}\mathrm{P}_W \mathrm{d}m_k \mathrm{d}z \mathrm{d}l
\end{align*}
}

Therefore 
\begin{align*}
&(2.2) + (2.4) + (2.5) - (2.8) = (2.9) + (2.10) \\ & +\int \left( \frac{\mathrm{g}\left(a^\prime\right)}{\mathrm{g}_1\left(a^\prime\right)} \frac{\mathrm{r}(m_k \mid a^\prime, z)}{\mathrm{r}_1(m_k \mid a^\prime, z)}  u(a,m_k)  -  u_1(a,m_k)  \frac{\mathrm{g}\left(a^\star\right)}{\mathrm{g}_1\left(a^\star\right)} \right) \left( \mathrm{q}_1(m_k \mid a^\star)  -  \mathrm{q}(m_k \mid a^\star) \right) \mathrm{d}\mathrm{P}_W \mathrm{d}m_k \mathrm{d}z \mathrm{d}l 
\end{align*}

Combing above results, the second-order term can be written as 
\begin{align*} 
& \int D_{P_1} (o) \mathrm{d}P-\theta_k^{\prime} = \int \left( v_1\left(a^\prime\right) - v\left(a^\prime\right) \right) \left(1 - \frac{\mathrm{g}\left(a^\prime\right)}{\mathrm{g}_1\left(a^\prime\right)}\right) \mathrm{d}\mathrm{P}_W \mathrm{d}m_k \mathrm{d}z \mathrm{d}l \tag{2.11}\\
&+\int \frac{\mathrm{g}\left(a^\prime\right)}{\mathrm{g}_1\left(a^\prime\right)} \left( \frac{\mathrm{q}_1(m_k \mid a^\star)}{\mathrm{r}_1(m_k \mid a^\prime, z)} - \frac{\mathrm{q}(m_k \mid a^\star)}{\mathrm{r}(m_k \mid a^\prime, z)} \right) \left( \mathrm{b}(m_k, a^\prime, z, l) - \mathrm{b}_1(m_k, a^\prime, z, l) \right) \mathrm{p}(z \mid a^\prime) \mathrm{p}(l \mid a^\prime, z, m_k) \mathrm{r}(m_k \mid a^\prime, z) \mathrm{d}\mathrm{P}_W \mathrm{d}m_k \mathrm{d}z \mathrm{d}l \tag{2.12} \\
& + \int \frac{\mathrm{g}\left(a^\prime\right)}{\mathrm{g}_1\left(a^\prime\right)} (s(a^\prime, z) - s_1(a^\prime, z)) \left( \frac{\mathrm{r}(m_k \mid a^\prime, z)}{\mathrm{r}_1(m_k \mid a^\prime, z)} - 1\right) \mathrm{p}(z \mid a^\prime)  \mathrm{d}\mathrm{P}_W \mathrm{d}m_k \mathrm{d}z \mathrm{d}l \tag{2.13} \\
& + \int \frac{\mathrm{g}\left(a^\prime\right)}{\mathrm{g}_1\left(a^\prime\right)}  \left( \mathrm{b}_1(m_k, a^\prime, z, l) - \mathrm{b}(m_k, a^\prime, z, l) \right)  \left( \frac{\mathrm{r}(m_k \mid a^\prime, z)}{\mathrm{r}_1(m_k \mid a^\prime, z)} - 1 \right) \mathrm{q}_1(m_k \mid a^\star) \mathrm{p}(z \mid a^\prime) \mathrm{p}(l \mid a^\prime, z, m_k) \mathrm{d}\mathrm{P}_W \mathrm{d}m_k \mathrm{d}z \mathrm{d}l \tag{2.14} \\
& + \int \frac{\mathrm{g}\left(a^\prime\right)}{\mathrm{g}_1\left(a^\prime\right)}  \left( \mathrm{b}_1(m_k, a^\prime, z, l) - \mathrm{b}(m_k, a^\prime, z, l) \right)  \left( \mathrm{q}_1(m_k \mid a^\star) -  \mathrm{q}(m_k \mid a^\star)  \right) \mathrm{p}(z \mid a^\prime) \mathrm{p}(l \mid a^\prime, z, m_k) \mathrm{d}\mathrm{P}_W \mathrm{d}m_k \mathrm{d}z \mathrm{d}l \tag{2.15}\\
& +\int \left( \frac{\mathrm{g}\left(a^\prime\right)}{\mathrm{g}_1\left(a^\prime\right)} \frac{\mathrm{r}(m_k \mid a^\prime, z)}{\mathrm{r}_1(m_k \mid a^\prime, z)}   -  1 \right) \left( \mathrm{q}_1(m_k \mid a^\star)  -  \mathrm{q}(m_k \mid a^\star)  \right) u(a,m_k) \mathrm{d}\mathrm{P}_W \mathrm{d}m_k \mathrm{d}z \mathrm{d}l \tag{2.16} \\
& + \int \left( u(a,m_k)   -  u_1(a,m_k) \right) \left( \mathrm{q}_1(m_k \mid a^\star)  -  \mathrm{q}(m_k \mid a^\star) \right) \mathrm{d}\mathrm{P}_W \mathrm{d}m_k \mathrm{d}z \mathrm{d}l \tag{2.17} \\
& + \int \left(  1 -  \frac{\mathrm{g}\left(a^\star\right)}{\mathrm{g}_1\left(a^\star\right)} \right) \left( \mathrm{q}_1(m_k \mid a^\star)  -  \mathrm{q}(m_k \mid a^\star)   \right) u_1(a,m_k)\mathrm{d}\mathrm{P}_W \mathrm{d}m_k \mathrm{d}z \mathrm{d}l \tag{2.18} 
\end{align*}

We can then derive the results for Lemma 1 in the main manuscript from expressions (2.11)--(2.18).
\end{proof}

\section{Proof of Theorem 4.2}
\begin{proof}
    Let $P_n$ denote the empirical distribution function of the observed data and $P_{n}^{\{j\}}$ denote the empirical distribution function for fold $\mathcal{U}_j$. Using the von Mises expansion, the cross-fitted one-step estimator can be expanded as
\begin{align*}
   \sqrt{n} (\hat{\theta}_{k,\text{cf-os}}^\prime - \theta_k^\prime) = \sqrt{n}(P_n-P)D_P + \frac{\sqrt{n}}{J} \sum_{j=1}^J (P_{n}^{\{j\}} - P) \left( D_{\hat{P}}^{\{j\}} - D_{P} \right) + \frac{\sqrt{n}}{J} \sum_{j=1}^J R(P, \hat{P}^{\{ j\} })
\end{align*}
where $D_P$ is the efficient influence function evaluated at the true distribution $P$, and $D_{\hat{P}}^{{j}}$ represents the efficient influence function evaluated at the estimated distribution for the $j$-th fold.

By the Central Limit Theorem, the first term converges to a normal, mean zero variable, so we have
 $$\sqrt{n}(P_n-P)D_P = o_P(1).$$ 
 
The second term, which is the empirical process term,
$$\frac{\sqrt{n}}{J} \sum_{j=1}^J (P_{n}^{\{j\}} - P) \left( D_{\hat{P}}^{\{j\}} - D_{P} \right) = o_P(1)$$ by the use of cross-fitting \citep{hines2022demystifying, Díaz2021nonpara}. 
The third term, which is the second-order remainder,
 $\frac{\sqrt{n}}{J} \sum_{j=1}^J R(P, \hat{P}^{\{ j\} }),$ can be shown to be $o_P(1)$ by applying the Cauchy-Schwarz inequality and Assumptions 1 and 2 to expressions (2.11)--(2.18). 
\end{proof}

\section{Partial TML estimator for \texorpdfstring{$\theta_k^\prime$ targeting $D_{P,Y}(o)$ and $D_{P,M_k}(o)$}{theta\_k\_prime}}

We propose a partial TML estimator for the components $D_{P,Y}(o)$ and $D_{P,M_k}(o)$ in the EIF, focusing specifically on binary $M_k$. Under the partial TML procedure, the sample averages of $D_{P,Y}(o)$ and $D_{P,M_k}(o)$ are set to zero, while the remaining EIF components are computed using the one-step estimator. 

In our example, the outcome $Y$ is binary. However, if $Y$ were continuous, the outcome should be transformed with values in $[0,1]$, and we can then use the transformed outcome in the TML estimator \citep{zheng2011cross}. Here, the TML procedures for $D_{P,Y}(o)$ and $D_{P,M_k}(o)$ follow the methodology described by \cite{Díaz2021nonpara}. Our cross-fitted partial TML estimator can be implemented using the following steps:

\begin{enumerate}
    \item Obtain initial cross-fitted estimates of the nuisance parameters, $\hat{\mathrm{b}}(m_k, a^\prime, z, l, w)$ and $\hat{\mathrm{q}}(1 \mid a^\star, w)$, and use these as initial estimates for the TML procedure.
    
    \item Following \cite{Díaz2021nonpara}, update the initial estimates $\hat{\mathrm{b}}(m_k, a^\prime, z, l, w)$ and $\hat{\mathrm{q}}(1 \mid a^\star, w)$ using logistic regression with covariates $H_Y$ and $H_{M_k}$, respectively, where
    \begin{equation*}
        \begin{aligned}
            H_Y &= \frac{\hat{\mathrm{q}}\left(m_k \mid a^\star, w\right)}{\hat{\mathrm{g}}\left(a^\prime \mid w\right) \hat{\mathrm{r}}(m_k \mid a^\prime, z, w)} \\
            H_{M_k} &= \frac{u(a^\prime, 1, w) - u(a^\prime, 0, w)}{\hat{\mathrm{g}}\left(a^\star \mid w\right)}.
        \end{aligned}
    \end{equation*}
    Let $\hat{\epsilon}_Y$ and $\hat{\epsilon}_{M_k}$ denote the estimated regression coefficients. Specifically, in practice, $\hat{\epsilon}_Y$ can be obtained by fitting a logistic regression with outcome $Y$, single covariate $H_Y$ and offset $\mathrm{logit}\left(\hat{\mathrm{b}}(m_k, a^\prime, z, l, w)\right)$ to the subset of data with $A = a^\prime$. Similarly, $\hat{\epsilon}_{M_k}$ can be obtained by fitting a logistic regression with outcome $M_k$, single covariate $H_{M_k}$ and offset $\mathrm{logit}\left(\hat{\mathrm{q}}(1 \mid a^\star, w)\right)$ to the subset of data with $A = a^\star$.
    Then, the updated estimates $\hat{\mathrm{b}}^\star(m_k, a^\prime, z, l, w)$ and $\hat{\mathrm{q}}^\star(1 \mid a^\star, w)$ are obtained as follows:
    \begin{equation*}
        \begin{aligned}
            \text{logit}\left(\hat{\mathrm{b}}^\star(m_k, a^\prime, z, l, w)\right) &= \text{logit}\left(\hat{\mathrm{b}}(m_k, a^\prime, z, l, w)\right) + \hat{\epsilon}_Y H_Y \\
            \text{logit}\left(\hat{\mathrm{q}}^\star(1 \mid a^\star, w)\right) &= \text{logit}\left(\hat{\mathrm{q}}(1 \mid a^\star, w)\right) + \hat{\epsilon}_{M_k} H_{M_k}.
        \end{aligned}
    \end{equation*}
    where $\mathrm{logit}(p) = \mathrm{log}(p/(1-p))$. 
    \item Replace $\hat{\mathrm{b}}(m_k, a^\prime, z, l, w)$ and $\hat{\mathrm{q}}(1 \mid a^\star, w)$ with $\hat{\mathrm{b}}^\star(m_k, a^\prime, z, l, w)$ and $\hat{\mathrm{q}}^\star(1 \mid a^\star, w)$, respectively and repeat Step 2 until the algorithm converges. The remaining components of the EIF, which are not targeted by the TML procedure, are calculated in the same manner as in the one-step estimator.
\end{enumerate}

\section{Efficient estimation for \texorpdfstring{$\theta_k$}{theta\_k} and \texorpdfstring{$\theta_{all}$}{theta\_{all}}}

\begin{theorem}
    (Efficient influence function for $\theta_k$) Since $L$ is absent from $\theta_k$, we redefine $u$ and $v$. For fixed $a^\star$ and $a^\prime$, we define 
\begin{align*}
v(a, w) &= \int \mathrm{b}\left(a, z, m_k, w\right) \mathrm{q}(m_k \mid a^\star, w) \mathrm{p}(z \mid a, w) \mathrm{d} m_k z \\
&= E \left\{ \int \mathrm{b}\left(a^\prime, z, m_k, w\right) \mathrm{q}(m_k \mid a^\star, w) \mathrm{d} m_k \mid A=a, W=w \right\}.
\end{align*}
\begin{align*}
u(m_k, a, w) &= \int \mathrm{b}(a, z, m_k, w) \mathrm{p}(z \mid a, w) \mathrm{d}z \\
&= E \left\{ \int \mathrm{b}(a, z, m_k, w) \mathrm{e}(a, z, m_k, w) \mid M_k=m_k, A=a, W=w \right\},
\end{align*}
where
\begin{align*}
\mathrm{e}(a, z, m_k, c) = \frac{\mathrm{q}(m_k \mid a, w)}{\mathrm{r}(m_k \mid a, z, w)}.
\end{align*}

The efficient influence function $D_{P}(o)$ for $\theta_k$ in a nonparametric model is
\begin{equation*}
\begin{aligned}
D_P(o) &= \frac{\mathbb{1}{\{a = a^\prime\}}}{\mathrm{g}\left(a^\prime \mid w\right)} \frac{\mathrm{q}(m_k\mid a^\star,w)}{\mathrm{r}(m_k \mid a^\prime, z,w)} \left\{Y - \mathrm{b}\left(a^\prime, z, m_k, w\right)\right\} \\
& + \frac{\mathbb{1}{\{ a =a^\prime} \}}{\mathrm{g}\left(a^\prime \mid w\right)} \left\{ \int \mathrm{b}\left(a^\prime, z, m_k, w\right) \mathrm{q}(m_k \mid a^\star, w) \mathrm{d}m_k - v(a^\prime, w) \right\} \\
& + \frac{\mathbb{1}{ \{a = a^\star \}}}{\mathrm{g}\left(a^\star \mid w\right)} \left\{u(m_k, a^\prime, w) - \int u(m_k, a^\prime, w) \mathrm{q}(m_k \mid a^\star, w) \mathrm{d}m_k \right\} \\
& + v\left(a^\prime, w\right) - \theta_k
\end{aligned}
\end{equation*} \label{eif_k}
\end{theorem}

\begin{theorem}
(Efficient influence function for $\theta_{all}$)  $Z$ is not included in $\theta_{all}$ compared to $\theta_k$, the efficient influence function $D_{P}(o)$ for $\theta_{all}$ in a nonparametric model is
\begin{align*}
D_P(o) &= \frac{\mathbb{1}{\{a=a^\prime\}}}{\mathrm{g}(a^\prime \mid w)} \frac{\mathrm{p}\left(a^\star \mid m, w\right)}{\mathrm{p}(a^\prime \mid m, w)} \{Y - \mathrm{b}(a^\prime, m, w)\} \\
& + \frac{\mathbb{1}{\{ a=a^\star \}}}{\mathrm{g}\left(a^\star \mid w\right)} \left(\mathrm{b}(a^\prime, m, w) - u(a^\star, w)\right) \\
& + u(a^\star, w) - \theta_{all},
\end{align*}
where
$$
u(a,w) = \int \mathrm{b}(a^\prime,m,w) \mathrm{q}\left(m \mid a, w\right) \mathrm{d} m  = E\left\{\mathrm{b}(a^\prime,m,w) \mid A=a,W=w \right\}.
$$\label{eif_all}
\end{theorem} 

According to Theorem \ref{eif_k} and \ref{eif_all},
the cross-fitted one-step estimator for $\theta_k$ can be derived in a similar way to that of $\theta_k^\prime$, following Steps 1-4 in Section 4.2.1 in the main manuscript, with adjustments for the exclusion of $L$. For the TML estimator, since $L$ is not involved, the full TML estimator for $\theta_k$ can be derived using the methods proposed by \cite{Díaz2021nonpara}, as $\theta_k$ is equivalent to the estimand described in \cite{Díaz2021nonpara} for the setting with high-dimensional exposure-induced mediator-outcome confounders. Efficient estimators for $\theta_{all}$ can be derived similarly to those for $\theta_k$, with adjustments for the exclusion of $Z$ further. The variance of these estimators is estimated using the sample variance of the EIF.
\newpage
\section{Metabolites list}
\begin{table}[H]
\begin{tabular}{l|l}
\hline
18:2, linoleic acid (mmol/L) & Mean diameter for VLDL particles (nm)\\

22:6, docosahexaenoic acid (mmol/L) & Phenylalanine (mmol/L) \\

Acetate (mmol/L)$^\star$ & Pyruvate (mmol/L)\\

Acetoacetate (mmol/L)$^\star$ & Ratio of apolipoprotein B to apolipoprotein AI\\

Estimated degree of unsaturation & Ratio of triglycerides to phosphoglycerides\\

Monounsat. fatty acids; 16:1, 18:1 (mmol/L) & Remnant cholesterol (nonHDL, nonLDL cholesterol) (mmol/L)\\

Omega3 fatty acids (mmol/L) & Serum total cholesterol (mmol/L)\\

Omega6 fatty acids (mmol/L) & Serum total triglycerides (mmol/L)$^\star$\\

Phosphatidylcholine \& other cholines (mmol/L) & Sphingomyelins (mmol/L)\\

Polyunsat. fatty acids (mmol/L) & Total cholesterol in HDL (mmol/L)\\

Ratio of 18:2 linoleic acid to total fatty acids (\%) & Total cholesterol in HDL2 (mmol/L)\\

Ratio of 22:6 docosahexaenoic acid to total fatty acids (\%) & Total cholesterol in HDL3 (mmol/L)\\

Ratio of monounsat. fatty acids to total fatty acids (\%) & Total cholesterol in IDL (mmol/L)\\

Ratio of omega3 fatty acids to total fatty acids (\%) & Total cholesterol in LDL (mmol/L)\\

Ratio of omega6 fatty acids to total fatty acids (\%) & Total cholesterol in VLDL (mmol/L)\\

Ratio of polyunsat. fatty acids to total fatty acids (\%) & Total lipids in chylomicrons \& ex.large VLDL (mmol/L)$^\star$\\

Ratio of saturated fatty acids to total fatty acids (\%) & Total lipids in IDL (mmol/L)\\

Saturated fatty acids (mmol/L) & Total lipids in large HDL (mmol/L)\\

Total cholines (mmol/L) & Total lipids in large LDL (mmol/L)\\

Total fatty acids (mmol/L) & Total lipids in large VLDL (mmol/L)$^\star$\\

Total phosphoglycerides (mmol/L) & Total lipids in medium HDL (mmol/L)\\

\textbf{Glycoprotein acetyls, mainly a1acid glycoprotein (mmol/L)} & Total lipids in medium LDL (mmol/L)\\

Albumin (signal area) & Total lipids in medium VLDL (mmol/L) \\

Apolipoprotein A1 (g/L) & Total lipids in small HDL (mmol/L)\\

Apolipoprotein B (g/L) & Total lipids in small LDL (mmol/L)\\

Creatinine (mmol/L) & Total lipids in small VLDL (mmol/L)\\

Esterified cholesterol (mmol/L) & Total lipids in very large HDL (mmol/L)\\

Free cholesterol (mmol/L) & Total lipids in very large VLDL (mmol/L)$^\star$\\

Glutamine (mmol/L) & Total lipids in very small VLDL (mmol/L) \\

Glycine (mmol/L) & Triglycerides in HDL (mmol/L) \\

Histidine (mmol/L) & Triglycerides in IDL (mmol/L) \\

Isoleucine (mmol/L) & Triglycerides in LDL (mmol/L)\\

Leucine (mmol/L) & Triglycerides in VLDL (mmol/L)$^\star$ \\

Mean diameter for HDL particles (nm) & Tyrosine (mmol/L) \\

Mean diameter for LDL particles (nm) & Valine (mmol/L) \\
\hline
\end{tabular}
\caption{List of metabolites in the LSAC example. Metabolites marked with $\star$ have been log-transformed due to skewness. Metabolites listed before and after Glycoprotein acetyls (GlycA) are considered as causal ancestors and descendants of GlycA, respectively when estimating $\theta_k^\prime$.}
\end{table}

\newpage

\begin{table}[H]
    \centering
\begin{tabular}{llllrrrr}
\toprule
folds & estimator & cut-off & estimand & IIE & SE & CIlow & CIupp\\
\midrule
10 & One-step & 0.5 & $\theta_{all}$ & 0.074 & 0.029 & 0.018 & 0.131\\

 &  &  & $\theta_{k}$ & 0.036 & 0.018 & 0.002 & 0.071\\

 &  &  & $\theta_{k}^\prime$ & 0.064 & 0.018 & 0.029 & 0.098\\

 &  & 0.75 & $\theta_{all}$ & 0.069 & 0.028 & 0.014 & 0.125\\

 &  &  & $\theta_{k}$ & 0.002 & 0.036 & -0.069 & 0.073\\

 &  &  & $\theta_{k}^\prime$ & 0.045 & 0.035 & -0.024 & 0.115\\

 & TML & 0.5 & $\theta_{all}$ & 0.073 & 0.030 & 0.014 & 0.131\\

 &  &  & $\theta_{k}$ & 0.025 & 0.018 & -0.009 & 0.060\\

 &  &  & $\theta_{k}^\prime$ & 0.068 & 0.018 & 0.033 & 0.104\\

 &  & 0.75 & $\theta_{all}$ & 0.069 & 0.027 & 0.017 & 0.121\\

 &  &  & $\theta_{k}$ & -0.008 & 0.038 & -0.082 & 0.067\\

 &  &  & $\theta_{k}^\prime$ & 0.054 & 0.033 & -0.011 & 0.118\\
\cmidrule{1-8}
5 & One-step & 0.5 & $\theta_{all}$ & 0.071 & 0.027 & 0.019 & 0.123\\

 &  &  & $\theta_{k}$ & 0.032 & 0.025 & -0.017 & 0.081\\

 &  &  & $\theta_{k}^\prime$ & 0.065 & 0.016 & 0.033 & 0.096\\

 &  & 0.75 & $\theta_{all}$ & 0.071 & 0.024 & 0.023 & 0.119\\

 &  &  & $\theta_{k}$ & 0.002 & 0.044 & -0.083 & 0.088\\

 &  &  & $\theta_{k}^\prime$ & 0.059 & 0.037 & -0.013 & 0.131\\

 & TML & 0.5 & $\theta_{all}$ & 0.066 & 0.027 & 0.014 & 0.119\\

 &  &  & $\theta_{k}$ & 0.017 & 0.024 & -0.031 & 0.064\\

 &  &  & $\theta_{k}^\prime$ & 0.065 & 0.017 & 0.032 & 0.097\\

 &  & 0.75 & $\theta_{all}$ & 0.066 & 0.026 & 0.015 & 0.118\\

 &  &  & $\theta_{k}$ & -0.012 & 0.042 & -0.094 & 0.070\\

 &  &  & $\theta_{k}^\prime$ & 0.052 & 0.039 & -0.024 & 0.129\\
\bottomrule
\end{tabular}
\caption{Results for the LSAC example: Estimated interventional indirect effects for the estimands $\text{IIE}_{all}$, $\text{IIE}_k$, and $\text{IIE}_k^\prime$ with their standard errors (SE) and 95\% confidence intervals (CIlow, CIupp). Results are provided for one-step and TML estimators with cross-fitting, with different folds and cut-off values for GlycA. The TML estimates for $\text{IIE}_k^\prime$ are obtained using the partial TML estimator described in Section 4.}

\end{table}

\bibliographystyle{abbrvnat}
\bibliography{references}  %%% Remove comment to use the external .bib file (using bibtex).

\end{document}